\documentclass[useAMS,usenatbib]{mn2e}

\usepackage{times,graphicx,amssymb,natbib}

\newcommand{\degrees}{^{\circ}}
\newcommand{\msol}{M_{\rm \odot}}
\newcommand{\mjup}{M_{\rm Jup}}
\newcommand{\mearth}{M_{\rm \oplus}}

\title[Dynamical Forcing of Exomoon Climates]{Dynamical Effects on the Habitable Zone for Earth-like Exomoons}
\author[Duncan Forgan and David Kipping]{Duncan Forgan $^{1}$\thanks{E-mail:dhf@roe.ac.uk} and David Kipping$^{2,3}$ \\
$^{1}$Scottish Universities Physics Alliance (SUPA), Institute for Astronomy, University of Edinburgh, Blackford Hill, Edinburgh, EH9 3HJ, Scotland, UK \\
$^{2}$Harvard-Smithsonian Centre for Astrophysics, Cambridge, Massachusetts 02138, USA \\
$^{3}$Carl Sagan Fellow}

\begin{document}


\pagerange{\pageref{firstpage}--\pageref{lastpage}} \pubyear{}

\maketitle

\label{firstpage}

\begin{abstract}

\noindent With the detection of extrasolar moons (exomoons) on the horizon, it is important to consider their potential for habitability.  If we consider the circumstellar Habitable Zone (HZ, often described in terms of planet semi-major axis and orbital eccentricity), we can ask, ``How does the HZ for an Earth-like exomoon differ from the HZ for an Earth-like exoplanet?'' 

For the first time, we use 1D latitudinal energy balance modelling to address this question.  The model places an Earth-like exomoon in orbit around a Jupiter mass planet, which in turn orbits a Sun-like star.  The exomoon's surface is decomposed into latitudinal strips, and the temperature of each strip is evolved under the action of stellar insolation, atmospheric cooling, heat diffusion, eclipses of the star by the planet, and tidal heating.  We use this model to carry out two separate investigations.  

In the first, four test cases are run to investigate in detail the dependence of the exomoon climate on orbital direction, orbital inclination, and on the frequency of stellar eclipse by the host planet.  We find that lunar orbits which are retrograde to the planetary orbit exhibit greater climate variations than prograde orbits, with global mean temperatures around 0.1 K warmer due to the geometry of eclipses.  If eclipses become frequent relative to the atmospheric thermal inertia timescale, climate oscillations become extremely small.

In the second investigation, we carry out an extensive parameter study, running the model many times to study the habitability of the exomoon in the 4-dimensional space composed of the planet semi-major axis and eccentricity, and the moon semi-major axis and eccentricity.  We find that for zero moon eccentricity, frequent eclipses allow the moon to remain habitable in regions of high planet eccentricity, but tidal heating severely constrains habitability in the limit of high moon eccentricity, making the habitable zone a sensitive function of moon semi-major axis.

\end{abstract}

\begin{keywords}

astrobiology, methods:numerical, planets and satellites: general

\end{keywords}

\section{Introduction}


\noindent The increasing rate of discovery of extrasolar planets (exoplanets) has given astronomers licence to consider their potential as habitats for extraterrestrial life.  To date, more than 860 exoplanets have been confirmed\footnote{http://exoplanet.eu}, with over 2,300 exoplanet candidates identified by the Kepler mission \citep{Batalha2013}.  Of the confirmed exoplanets, several are in the circumstellar Habitable Zone (HZ), an annular region surrounding the star where planets with orbits inside it may be expected to possess liquid water on their surface.  

The HZ is derived for planets of terrestrial mass and composition - the inner edge is determined  by water loss (via hydrogen escape and photolysis), and the outer edge is determined by runaway glaciation due to a build up of $CO_2$ clouds \citep{Kasting_et_al_93}.  In the case of single stars, the HZ is a circular annulus, and hence terrestrial planets on circular orbits inside the annulus are expected to be habitable.  Even if the planet's orbit is not circular, it may still be habitable depending on the average received flux over its orbit, which can vary significantly if the orbit is highly elliptical \citep{Williams2002,Kane2012,Kane2012c}.  If the star is a multiple system, the geometry of the HZ annulus becomes time-dependent, and this can have important consequences for the climates of planets in the system \citep{Forgan2012,Eggl2012,Eggl2012a,Kane2013}.

To date, all exoplanets that reside within their local single-star HZ have radii 1.5 to 2 times larger than that of the Earth, and as such may be mini-Neptunes than super-Earths (see for example \citealt{Barnes2009}).  Even if such planets were terrestrial in nature, they are likely to have significantly higher mantle viscosity, which suppresses convection and as a result has implications for heat transfer and the planet's ability to degas \citep{Stamenkovic2010,Liu2010}.

This being the case, the current crop of exoplanets in the HZ should not be ruled out as a source of habitable niches.  Extrasolar moons (exomoons) belonging to these planets may be terrestrial in nature, and as a result may be habitable \citep{Williams1997b}.  This is true even in our own Solar System: outside the Earth, the best chances for life in the Solar System are believed to be on tidally heated icy moons which are thought to possess liquid subsurface oceans, such as Europa \citep{Melosh2004} and Enceladus \citep{plume_enceladus}. Titan, the largest moon in the Solar System, has a thick atmosphere, with evidence of precipitation and lakes constituting a hydrological cycle \citep{lake_titan}.

Exomoons are yet to be detected, but Earth-mass exomoons are in principle detectable indirectly with Kepler \citep{Kipping_moon}.  Studying an exoplanet's transit timing variations and transit duration variations (TTV and TDV respectively) allows observational constraints on the moon's mass, and its orbital semi-major axis around the host exoplanet. Additionally, auxiliary transits of the moon and planet-moon mutual events lead to unique transit features revealing the moon's radius and other orbital parameters \citep{Kipping2011}.  Such efforts to realise a detection are underway as part of the Hunt for Exomoons with Kepler (HEK) project \citep{Kipping2012}.

A large component of habitability studies is the study of the energy balance associated with the celestial body in question.  The various sources and sinks of heat are catalogued and compared, and the system is evolved over time to understand their interplay, either by analytical calculation or numerical simulation.  In the case of exomoons, the list of sources is longer than that of exoplanets.  Alongside direct stellar insolation, exomoons receive starlight reflected from the planet (as well as the planet's own radiation), and they are tidally heated.  Also, the moon is more prone to stellar eclipses as a result of the host planet, providing an extra (effective) energy sink.

The study of habitable exomoons has grown apace in recent years: early calculations by \citet{Reynolds1987} demonstrated that Europa is a viable niche for marine life such as that found under Antartic ice, and proposed a tidally heated circumplanetary habitable zone.  In the early phase of exoplanet detection, \citet{Williams1997b} suggested that two of the then 9 known giant exoplanets could host habitable moons.  They raised the three key obstacles to exomoon habitability, namely:

\begin{enumerate}
\item Tidal locking of the moon with respect to the planet, resulting in extreme weather conditions \citep{Joshi1997}.
\item Increased EUV and X-Ray radiation from the host planet's magnetosphere leading to loss of the moon's atmosphere \citep{Kaltenegger2000}, and
\item A poorly proportioned distribution of volatile abundances, due to the differing environments of planet and moon formation.  \citet{Kaltenegger2000} notes that Solar System moons may be   over-abundant with water to be habitable, if placed in the inner Solar System, as they would lack the large areas of land required to sustain a carbonate-silicate cycle.  Without this, the body would lack a crucial regulatory system to ameliorate the effect of the star's growing luminosity as it evolves along the Main Sequence.
\end{enumerate} 

\citet{Scharf2006} investigated the then-known exoplanet population for the potential to host habitable exomoons. Estimating that almost 30\% of the total population could host small moons with icy mantles, he goes on to estimate that tidal heating could increase the habitable zone's extent in planet semi-major axis by a factor of around 2 (although he notes that the tidal heating required can become several orders of magnitude larger than that experienced by Io, one of the more extreme cases in our Solar System).

\citet{Heller2013} points out that exomoons may in some cases provide better niches than exoplanets, for several reasons:

\begin{enumerate}
\item Exoplanets in the HZ of low mass stars (e.g. \citealt{Dressing2013}) are likely to be tidally locked, whereas moons of such exoplanets are likely to be tidally locked \emph{to the planet, not the star} \citep{Kaltenegger2010,Sasaki2012}.  This reduces the strong fluctuation of surface insolation (and potential atmospheric collapse) that a moon would suffer if it was tidally locked to the star \citep{Joshi1997, Kite2011}.  This may also provide energy to drive the moon's internal dynamo, and generate a magnetosphere to guard against atmospheric loss.
\item Exoplanet obliquity will erode due to tidal evolution, as demonstrated by the axial orientations of Venus and Mercury.  This erosion is resisted more easily by more massive planets, if its
  orbital semi major axis is not too low and the planet does not possess relatively massive neighbours \citep{Heller2011,Sasaki2012}.  As a result, exomoons tidally locked to their host planet can retain the obliquity of their host planet and not suffer such strong erosion.
\end{enumerate} 

In their work, \citet{Heller2013} calculate the time dependent flux upon a moon due to the star's insolation, reflected starlight and thermal emission from the planet's surface, and add to this the tidal heating received. They use this total effective flux to calculate a parameter space of habitability in the moon's orbit about the host planet, and the host planet's mass, which they investigate for Kepler-22b and KOI211.01, both of which reside in the circumstellar HZ.  By averaging over the moon's orbital period, they produce a corrective term for the circumstellar HZ depending on the planet's albedo and the moon's orbital semi-major axis relative to the planet, as well as surface temperature maps for the moons.  However, while these calculations are among the most sophisticated to date, they do not allow for the transfer of heat due to an atmosphere.  This is clearly a crucial component of any climate model, particularly in the case of exomoon systems, which present many opportunities to set up strong temperature gradients that will generate heat transport.

In this paper, we implement a 1D latitudinal energy balance model (LEBM) to study the climate of an Earth-like exomoon in orbit around a Jupiter mass planet, which in turn orbits a Sun-like star.  Our use of a LEBM allows us to go beyond analytical calculations of detailed energy balance by allowing the moon to transfer heat across its surface (via a simple diffusion approximation) as well as adding some of the sources and sinks of energy described above.  

Rather than investigate the circumstellar HZ separately from the circumplanetary habitable region, we choose instead to investigate them together as a combined four dimensional manifold, constituted by the planet's semi-major axis and eccentricity relative to the star ($a_p$, $e_p$) and the moon's semi-major axis and eccentricity relative to the planet ($a_m$, $e_m$).  We wish to investigate how the extra dimensions of this manifold produces features unique to exomoon habitability, and how other factors, such as the moon's orbital direction, affect the resulting climate.

We investigate this problem in two parts: in the first, we simulate four simple test cases to study in detail the effect of orbital dynamics on the resulting climate.  In the second, we carry out a large parameter study of simulations to investigate the four dimensional habitable manifold.

In section \ref{sec:LEBM} we describe the LEBM and its implementation; in section \ref{sec:results} we display results from our investigations; in section \ref{sec:discussion} we discuss the results, and in section \ref{sec:conclusions} we summarise the work.

\section{Latitudinal Energy Balance Modelling} \label{sec:LEBM}


\subsection{Simulation Setup}

\noindent In all simulations, we fix the star mass at $M_*=1 \msol$, the mass of the host planet at $M_p = 1 \mjup$, and the mass of the moon at 1$\mearth$.  This system has been demonstrated to be dynamically stable on timescales comparable to the Solar System lifetime \citep{Barnes2002}.  

\begin{figure*}
\begin{center}$
\begin{array}{cc}
\includegraphics[scale = 0.25]{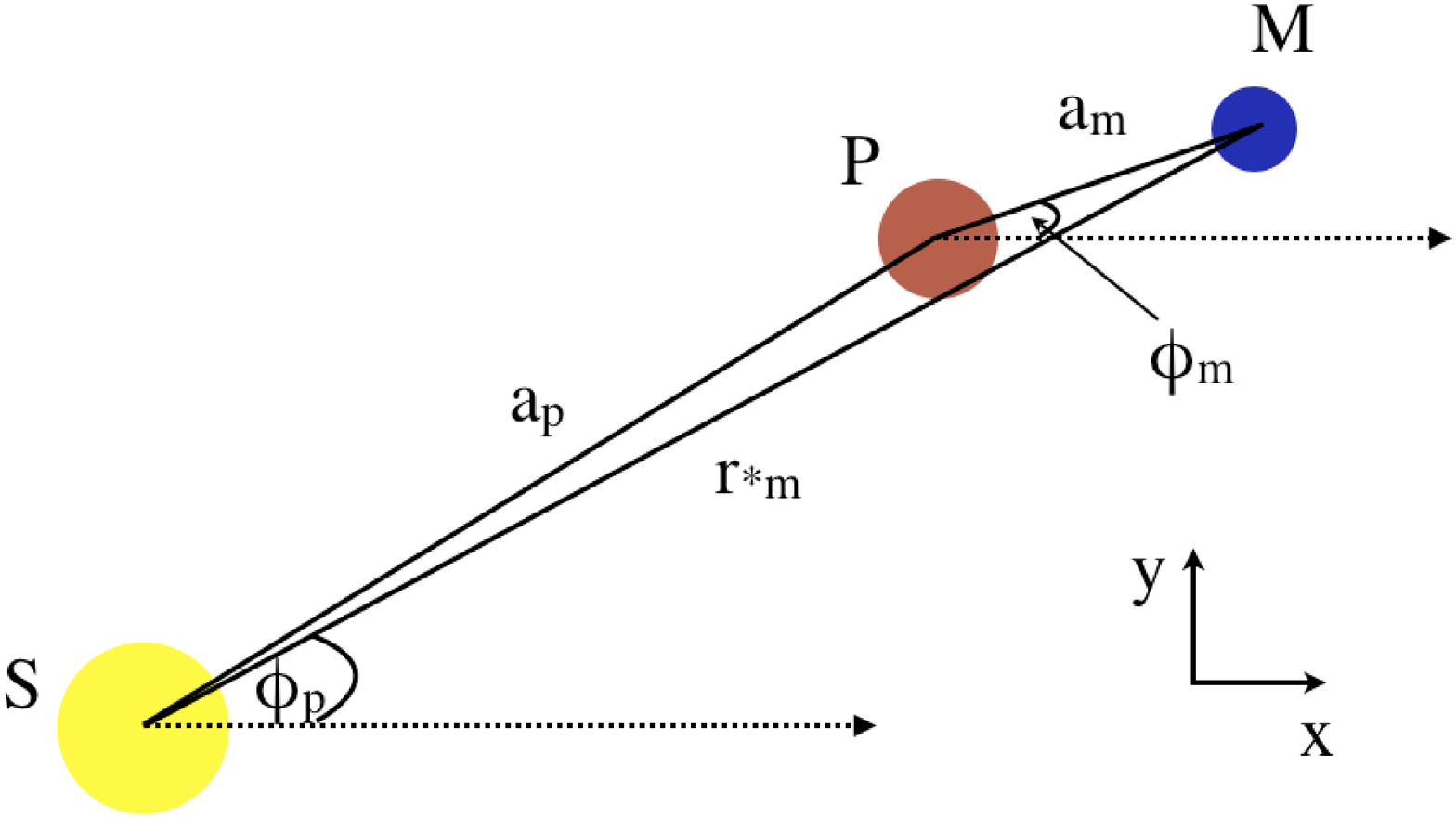} &
\includegraphics[scale=0.25]{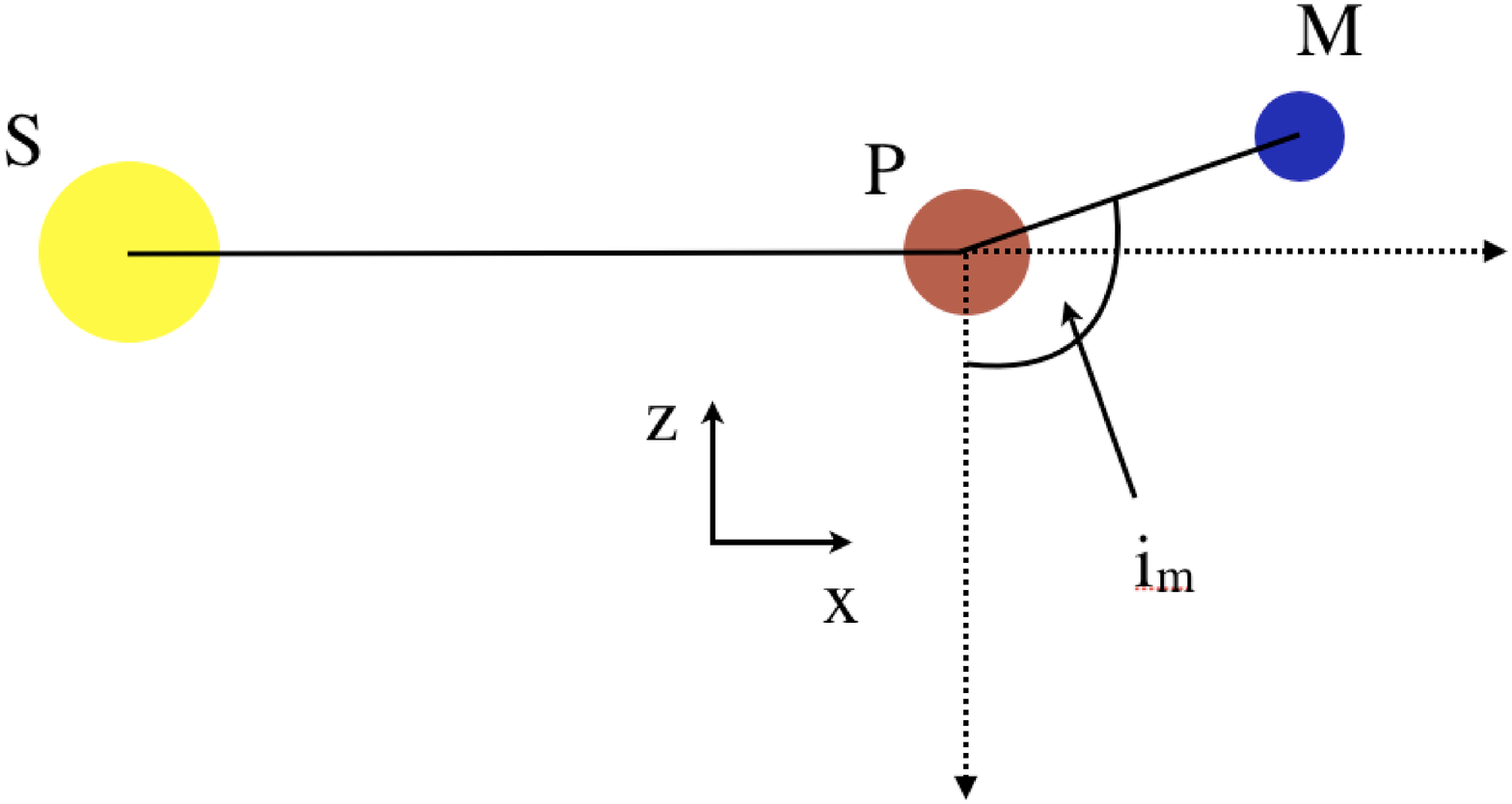}
\end{array}$
\caption{The setup for the latitudinal energy balance model (LEBM), shown in the x-y and x-z planes respectively.  We assume the moon's mass is negligible compared to the planet mass, and as such the moon orbits the planet, not the barycentre of the planet-moon system. \label{fig:setup}}
\end{center}
\end{figure*}

The planet's orbit around the star is specified by its semi-major axis $a_p$ and its eccentricity $e_p$, and the moon's orbit around the planet is specified by its semi-major axis $a_m$, eccentricity $e_m$ and inclination relative to the planet's orbital plane $i_m$ (see Figure \ref{fig:setup}).  The orbital longitudes of the planet and moon are defined such that $\phi_{p}=\phi_{m}=0$ corresponds to the x-axis.  Note that as the moon mass is much less than the planet mass, we assume $a_m$ is equal to the distance of the moon from the planet, as this is approximately equal to the distance from the moon to the barycentre of the planet-moon system.

Comparing this to the more traditional setup for exoplanet LEBMs, there are two principal differences:

\begin{enumerate}
\item The epicyclic motion of the moon relative to the star (as described by the distance between the star and the moon, $r_{*m}$), and
\item The potential for stellar eclipses by the planet. 
\end{enumerate}

\noindent A third difference is the potential for an extra energy source in the form of tidal heating, which we approximate using the expressions in \citet{Scharf2006}.  While our calculations are of a lower dimension than those of \citet{Heller2013}By allowing for diffusion of heat across latitudes, we also differ from \citet{Heller2013}.  We describe the equations that constitute the LEBM in the following section.

\subsection{A One Dimensional Latitudinal Energy Balance Model with Tidal Heating}

\noindent In this paper, we adapt the one dimensional LEBM applied to planets to function for moons with an atmosphere of similar composition to the Earth.  LEBMs solve the following diffusion equation:

\begin{equation} 
C \frac{\partial T}{\partial t} - \frac{\partial }{\partial x}\left(D(1-x^2)\frac{\partial T}{\partial x}\right) = S\left[1-A(T)\right] - I(T), 
\end{equation}

\noindent where $T=T(x,t)$ is the temperature at time $t$, $x = \sin \lambda$, and $\lambda$ is the latitude (between $-90\degrees$ and $90\degrees$).  This equation is evolved with the boundary condition $\frac{dT}{dx}=0$ at the poles.  The $(1-x^2)$ term is a geometric factor, arising from solving the diffusion equation in spherical geometry.

$C$ is the effective heat capacity of the atmosphere, $D$ is a diffusion coefficient that determines the efficiency of heat redistribution across latitudes, $S$ is the insolation received from the star, $I$ is the atmospheric infrared cooling and $A$ is the moon's albedo.  In the above equation, $C$, $S$, $I$ and $A$ are functions of $x$ (either explicitly, as $S$ is, or implicitly through $T(x,t)$).   

$D$ is defined such that an Earth-like planet at 1 AU around a star of $1 \msol$, with diurnal period of 1 day will reproduce the average temperature profile measured on Earth, see e.g. \citet{Spiegel_et_al_08}.  Planets that rotate rapidly experience inhibited latitudinal heat transport, due to Coriolis effects (see \citealt{Farrell1990}).  We follow \citet{Spiegel_et_al_08} by scaling $D$ according to:

\begin{equation} 
D=5.394 \times 10^2 \left(\frac{\omega_d}{\omega_{d,\oplus}}\right)^{-2},\label{eq:D}
\end{equation}

\noindent where $\omega_d$ is the rotational angular velocity of the planet, and $\omega_{d,\oplus}$ is the rotational angular velocity of the Earth.   This is a simplified expression, which does not describe the full effects of rotation, but allows for rapid computation without severely compromising the model's accuracy.

The diffusion equation is solved using a simple explicit forward time, centre space finite difference algorithm (as was done in \citealt{Forgan2012}).  A global timestep was adopted, with constraint

\begin{equation}
\delta t < \frac{\left(\Delta x\right)^2C}{2D(1-x^2)}.  
\end{equation}

The parameters are diurnally averaged, i.e. a key assumption of the model is that the moons rotate sufficiently quickly relative to their orbital period. This is clearly not applicable for certain orbital parameters, as tidal locking will play a significant role at low values of $a_m$ \citep{Joshi1997}.

The atmospheric heat capacity depends on what fraction of the planet's surface is ocean, $f_{ocean}$, what fraction is land $f_{land}=1.0-f_{ocean}$, and what fraction of the ocean is frozen $f_{ice}$:

\begin{equation} 
C = f_{land}C_{land} + f_{ocean}\left((1-f_{ice})C_{ocean} + f_{ice} C_{ice}\right). 
\end{equation}

\noindent The heat capacities of land, ocean and ice covered areas are 

\begin{equation} 
C_{land} = 5.25 \times 10^9 $ erg cm$^{-2}$ K$^{-1},
\end{equation}

\begin{equation} C_{ocean} = 40.0C_{land},\end{equation}
\begin{equation} C_{ice} = \left\{
\begin{array}{l l }
9.2C_{land} & \quad \mbox{263 K $< T <$ 273 K} \\
2C_{land} & \quad \mbox{$T<263$ K} \\
0.0 & \quad \mbox{$T>$ 273 K}.\\
\end{array} \right. 
\end{equation}

\noindent The infrared cooling function is 

\begin{equation} 
I(T) = \frac{\sigma_{SB}T^4}{1 +0.75 \tau_{IR}(T)}, 
\end{equation}

\noindent where the optical depth of the atmosphere 

\begin{equation} 
\tau_{IR}(T) = 0.79\left(\frac{T}{273\,\mathrm{K}}\right)^3. 
\end{equation}

\noindent The albedo function is

\begin{equation} 
A(T) = 0.525 - 0.245 \tanh \left[\frac{T-268\, \mathrm K}{5\, \mathrm K} \right]. 
\end{equation}

\noindent This produces a rapid shift from low albedo to high albedo as the temperature drops below the freezing point of water, producing highly reflective ice sheets. It is this transition that makes the outer habitable zone extremely sensitive to changes to various orbital and planetary parameters, and makes LEBMs an important tool in studying short-term temporal evolution of planetary climates.  

The insolation flux $S$ is a function of both season and latitude.  At any instant, the bolometric flux received at a given latitude at an orbital distance $r$ is

\begin{equation}
S = q_0\cos Z \left(\frac{1 AU}{r}\right)^2,
\end{equation}

\noindent where $q_0$ is the bolometric flux received from the star at a distance of 1 AU, and $Z$ is the zenith angle:

\begin{equation} 
q_0 = 1.36\times 10^6\left(\frac{M_*}{\msol}\right)^4 \mathrm{erg \,s^{-1}\, cm^{-2}} 
\end{equation}

\begin{equation} 
\cos Z = \mu = \sin \lambda \sin \delta + \cos \lambda \cos \delta \cos h. 
\end{equation} 

\noindent We have assumed a simple main sequence scaling for the luminosity. $\delta$ is the solar declination, and $h$ is the solar hour angle.  The solar declination is calculated from the obliquity $\delta_0$ as:

\begin{equation} 
\sin \delta = -\sin \delta_0 \cos(\phi_{*m}-\phi_{peri,m}-\phi_a), 
\end{equation}

\noindent where $\phi_{*m}$ is the current orbital longitude of the moon \emph{relative to the star}, $\phi_{peri,m}$ is the longitude of periastron, and $\phi_a$ is the longitude of winter solstice, relative to the longitude of periastron.   We set $\phi_{peri,m}=\phi_a=0$ for simplicity. 

We must diurnally average the solar flux:

\begin{equation} 
S = q_0 \bar{\mu}. 
\end{equation}

\noindent This means we must first integrate $\mu$ over the sunlit part of the day, i.e. $h=[-H, +H]$, where $H$ is the radian half-day length at a given latitude.  Multiplying by the factor $H/\pi$ (as $H=\pi$ if a latitude is illuminated for a full rotation) gives the total diurnal insolation as

\begin{equation} 
S = q_0 \left(\frac{H}{\pi}\right) \bar{\mu} = \frac{q_0}{\pi} \left(H \sin \lambda \sin \delta + \cos \lambda \cos \delta \sin H\right). \label{eq:insol}
\end{equation}

\noindent The radian half day length is calculated as

\begin{equation} 
\cos H = -\tan \lambda \tan \delta. 
\end{equation}

\noindent In the interest of computational expediency, we make a simple approximation for tidal heating, by firstly assuming the tidal heating per unit area is \citep{Peale1980,Scharf2006}:

\begin{equation} 
\dot{e}_T = \frac{21}{38} \frac{\rho^2_m R^{5/2}_m e^2_m}{\Gamma Q}\left(\frac{M_p}{a^3_m}\right)^{5/2} 
\end{equation}

\noindent where $\Gamma$ is the moon's elastic rigidity (which we assume to be uniform throughout the body), $R_m$ is the moon's radius, $\rho_m$ is the moon's density, $M_p$ is the planet mass, $a_m$ and $e_m$ are the moon's orbital semi-major axis and eccentricity (relative to the planet), and $Q$ is the moon's tidal dissipation parameter.  We assume terrestrial values for these parameters, hence $Q=100$, $\Gamma=10^{11} \,\mathrm{dyne \, cm^{-2}}$ (appropriate for silicate rock) and $\rho_m=5 \, \mathrm{g \, cm^{-3}}$.

To calculate how this heating is distributed latitudinally, we assume the heating rate is at its maximum at the subplanetary point.  We reuse equation (\ref{eq:insol}), substituting $q_0$ for $\dot{e}_T$, $\delta$ for the appropriate $\delta$ of the moon relative to the planet (which in this case is equal to zero as we assume $\delta_0$, i.e the relative obliquity between the moon and the planet, is zero), and fixing $H=\pi$ (i.e. tidal heating occurs throughout the diurnal period of the moon).  This is very much an approximation - indeed, the multi-dimensional nature of tidal heating prohibits latitudinal models from improving much on approximations such as this - habitability studies typically assume tidal heating is uniformly distributed throughout the body (see e.g. \citealt{Heller2013}).

\noindent We calculate habitability indices in the same manner as \citet{Spiegel_et_al_08}.  The habitability function $\xi$ is\footnote{This function is often labelled $\eta$ - we choose $\xi$ to avoid confusion with the use of $\eta$ as the frequency of Earth-like/habitable planets used in the exoplanets/SETI communities}:

\begin{equation} 
\xi(\lambda,t) = \left\{
\begin{array}{l l }
1 & \quad \mbox{273 K $< T(\lambda,t) <$ 373 K} \\
0 & \quad \mbox{otherwise}. \\
\end{array} \right. \end{equation}

\noindent We then average this over latitude to calculate the fraction of habitable surface at any timestep:

\begin{equation} 
\xi(t) = \frac{1}{2} \int_{-\pi/2}^{\pi/2}\xi(\lambda,t)\cos \lambda \, d\lambda. 
\end{equation}

\noindent Each simulation is allowed to evolve until it reaches a steady or quasi-steady state, and the final ten years of climate data are used to produce a time-averaged value of $\xi(t)$, $\bar{\xi}$, and the sample standard deviation, $\sigma_{\xi}$.  We use these two parameters to classify each simulations as follows:

\begin{enumerate}
\item \emph{Habitable Moons} - these moons possess a time-averaged $\bar{\xi}>0.1$, and $\sigma_{\xi} < 0.1\bar{\xi}$, i.e. the fluctuation in habitable surface is less than 10\% of the mean.
\item \emph{Hot Moons} - these moons have temperatures above 373 K across all seasons, and are therefore conventionally uninhabitable ($\bar{\xi} <0.1$).
\item \emph{Snowball Moons} - these moons have undergone a snowball transition to a state where the entire moon is frozen, and are therefore conventionally uninhabitable ($\bar{\xi}<0.1$).
\item \emph{Transient Moons} - these moons possess a time-averaged $\bar{\xi}>0.1$, and $\sigma_{\xi} > 0.1\bar{\xi}$, i.e. the fluctuation in habitable surface is greater than 10\% of the mean.
\end{enumerate}

\section{Results}\label{sec:results}

\noindent Exomoon systems present a parameter space of high dimension.  We therefore take two approaches to exploring subsets of this space.  The first approach fixes the planet's orbit, and considers four different sets of moon orbital parameters as test cases.

The second is a larger survey of parameter space, systematically varying $a_p,e_p,a_m,e_m$ (with the other orbital parameters held fixed).  This allows us to map out a four-dimensional manifold to represent the habitable zone, a subset of the true, higher dimensional habitable zone manifold (which will depend on extra parameters such as the obliquity of the moon and its chemical composition, which we do not vary).

\subsection{Four Test Cases}

\noindent We fix the orbit of the planet with $a_p= 0.9$ AU with $e_p=0$, and consider four simple test cases to probe the effects of epicylic motion and eclipses on exomoon habitability.  

\begin{enumerate}
\item A prograde circular orbit (i.e. where the orbital angular momenta of the planet and moon are aligned), with $a_m = 0.023 \mathrm{AU} = 481 R_p$ (approximately 0.3 Hill Radii), with $i_m=90\degrees$
\item As 1., but with a retrograde orbit (where the orbital angular momenta are anti-aligned, and $i_m=270\degrees$).
\item As 1., but with a semi-major axis reduced by a factor of 5: $a_m = 0.0046 \mathrm{AU} = 96 R_p$, (approximately 0.06 Hill Radii).
\item As 1., but with $i_m = 0\degrees$ (i.e. a polar orbit).
\end{enumerate}

\noindent These parameter choices are not entirely physically motivated - for example, it is unlikely that the orbit described in case 3 would remain stable on long time scales \citep{Weidner2010}, not to mention that synchronous rotation would become important to the moon's climate.  We choose these cases to illustrate more clearly the dynamical effects of changing the orbital parameters of the moon.  We select $a_p=0.9$ AU to guarantee that all four test cases are habitable (see following section).

\begin{figure*}
\begin{center}$
\begin{array}{cc}
\includegraphics[scale = 0.4]{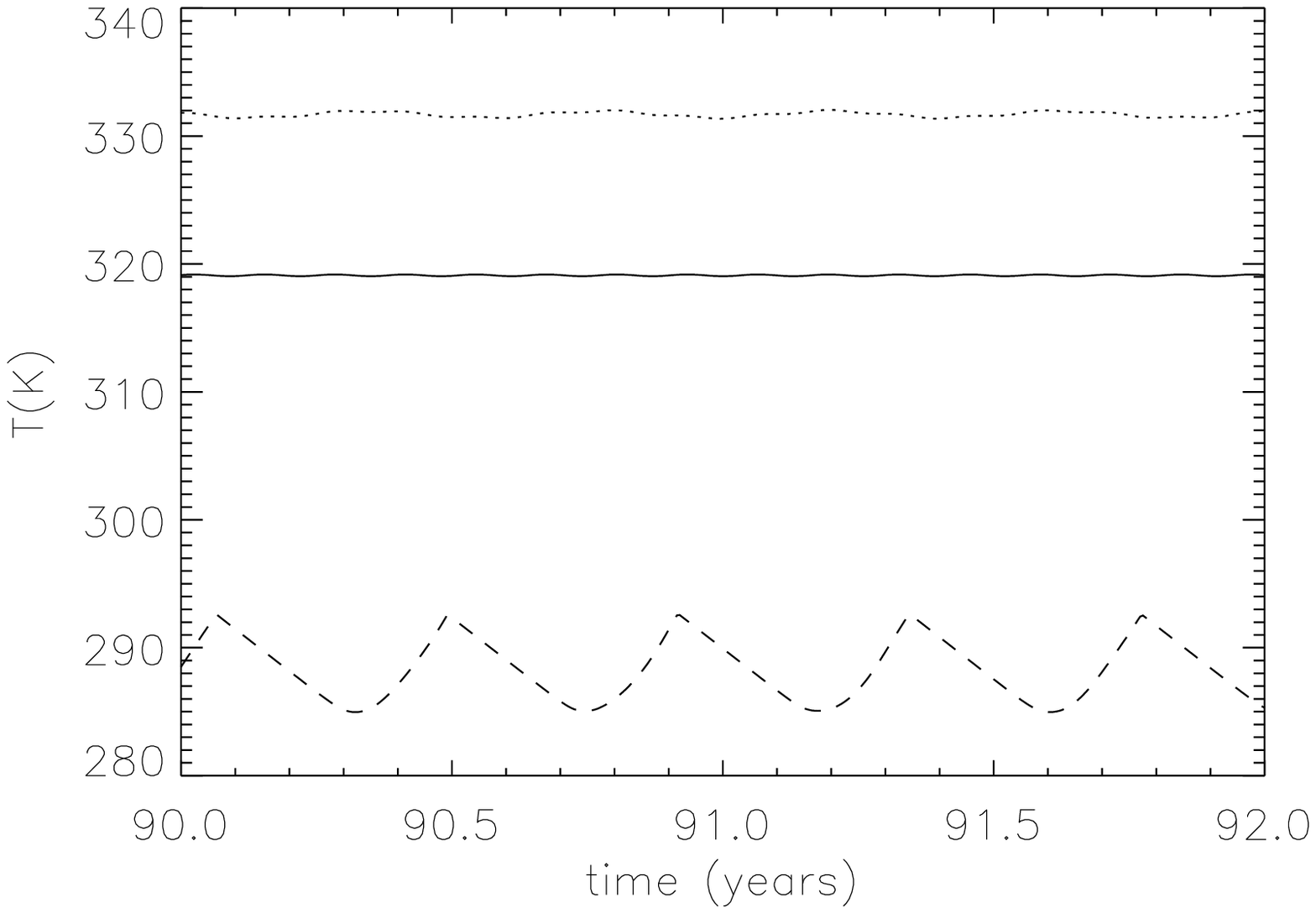} &
\includegraphics[scale=0.4]{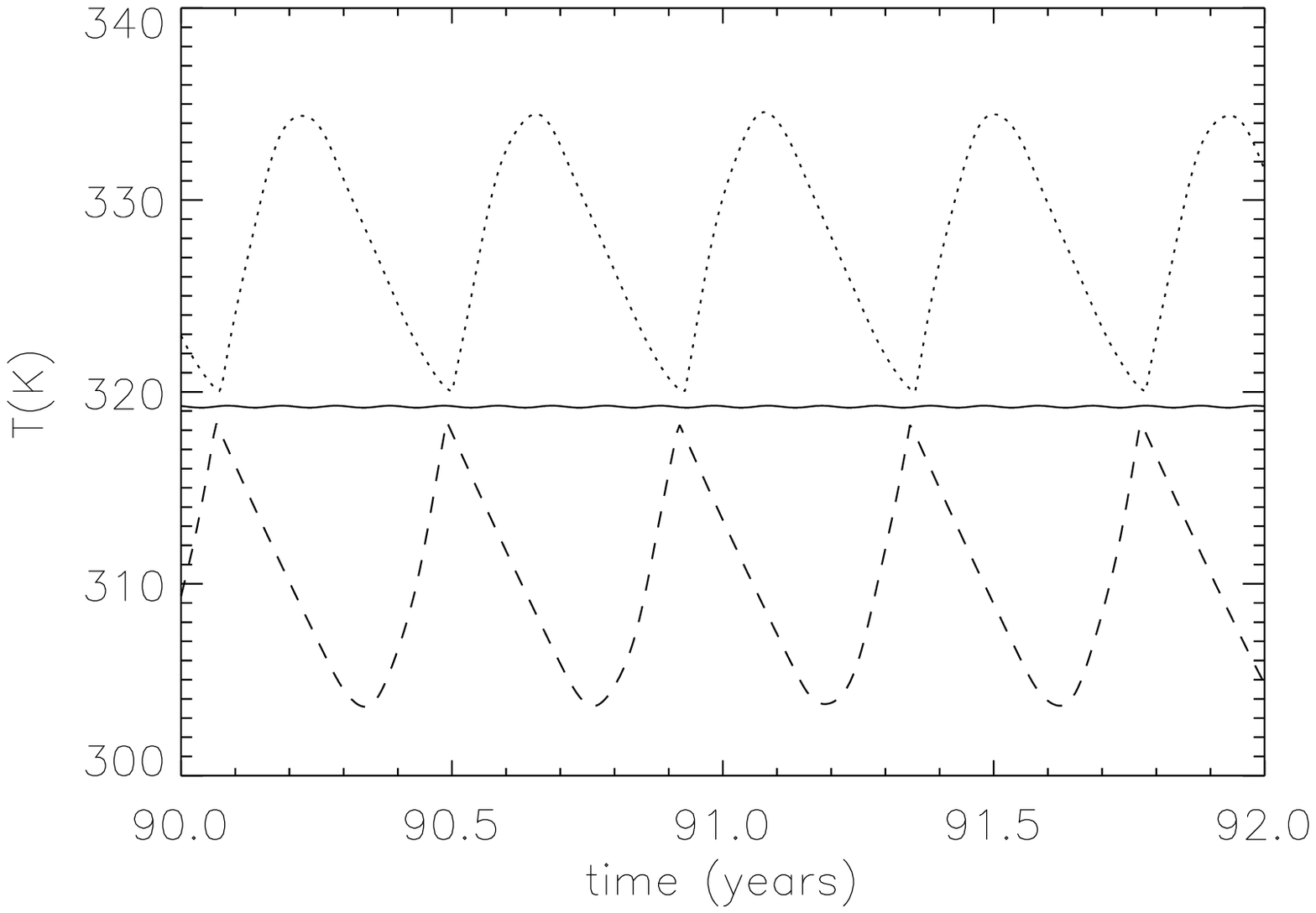} \\
\includegraphics[scale = 0.4]{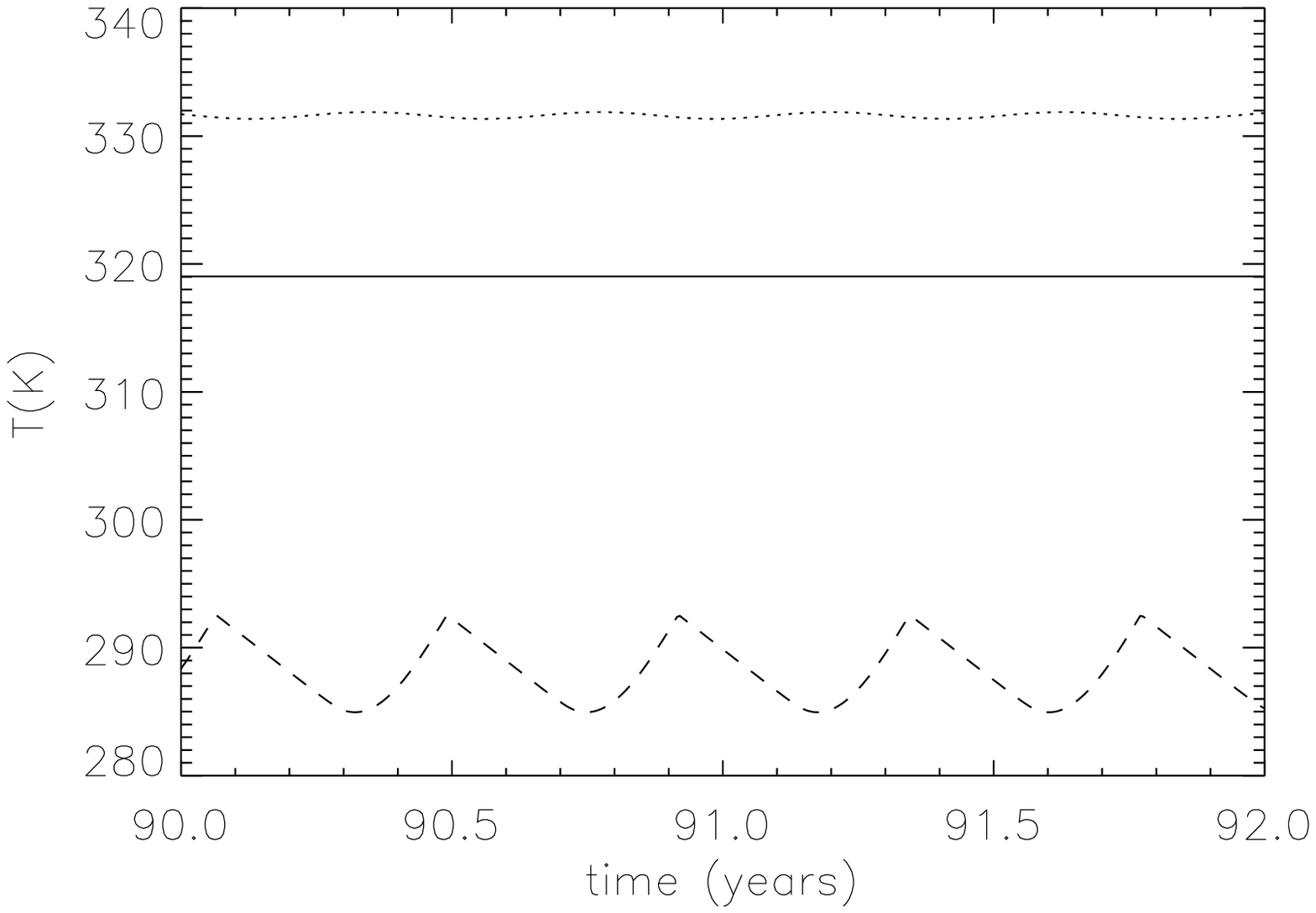} &
\includegraphics[scale=0.4]{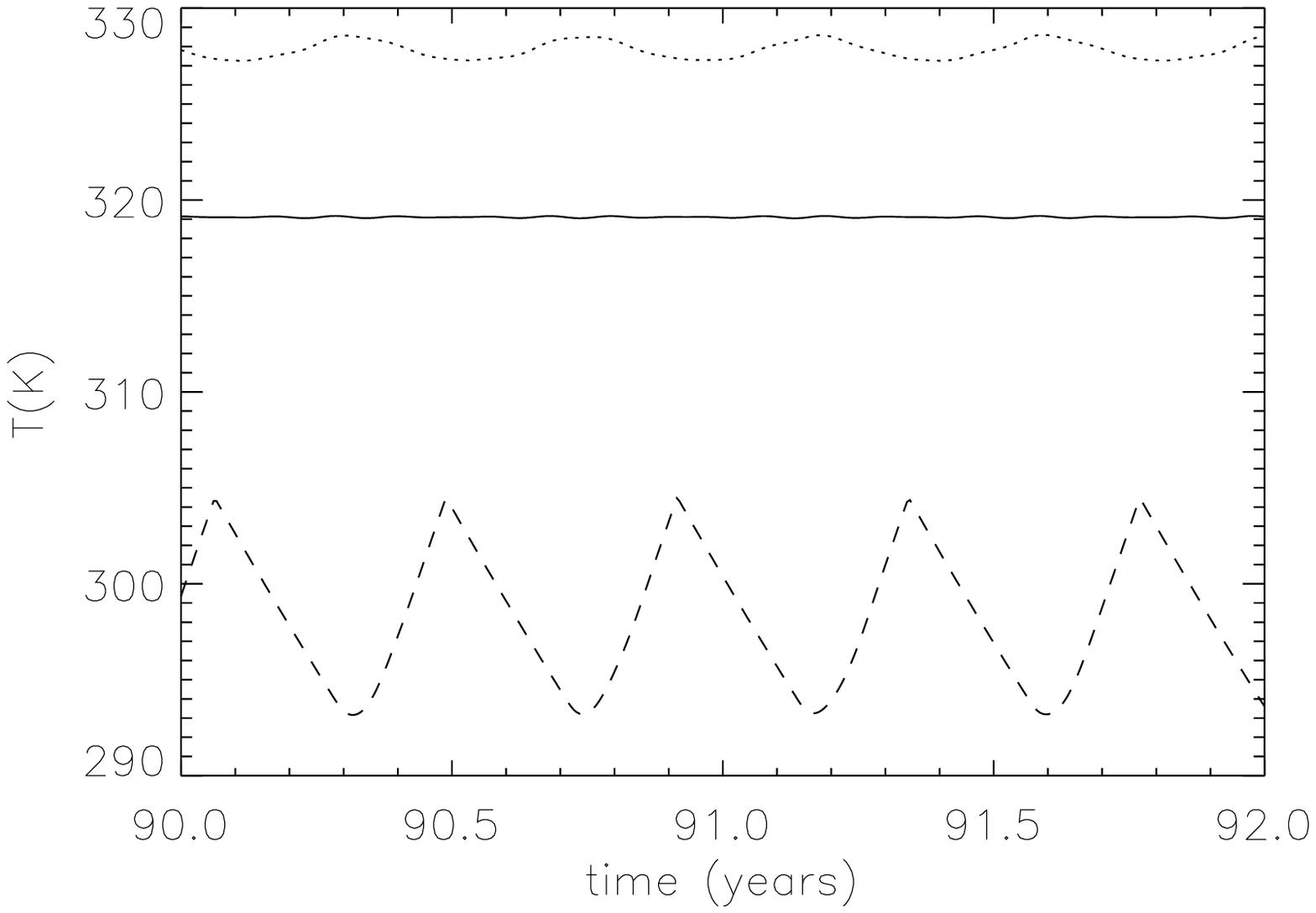} 
\end{array}$
\caption{The temperature evolution of the four test cases, over 2 years, approximately 18 moon orbital periods (560 for case 3), and approximately 2.35 planetary orbital periods).  The top left panel shows case 1, the top right panel shows case 2, the bottom left panel shows case 3, and the bottom right panel shows case 4.  The solid lines indicate the mean temperature, the dashed lines indicate the global minimum temperature, and the dotted lines indicate the global maximum temperature.   \label{fig:casescompareT}}
\end{center}
\end{figure*}

All four moons are entirely habitable ($\bar{\xi} = 1$).  This is easily established by studying the mean, minimum and maximum surface temperatures (Figure \ref{fig:casescompareT}).  The minimum temperature does not decrease beneath 290 K throughout the orbit of the moon and the planet.  The differences between prograde and retrograde moons are immediately obvious (top left and top right panel respectively).  While the mean temperatures appear to be similar, the maximum and minimum temperatures fluctuate much more strongly in the retrograde case.  Case 3 is very similar to case 1, and case 4 also has a similar mean temperature (with a more pronounced oscillation in the maximum and minimum temperatures).

\begin{figure*}
\begin{center}$
\begin{array}{cc}
\includegraphics[scale = 0.4]{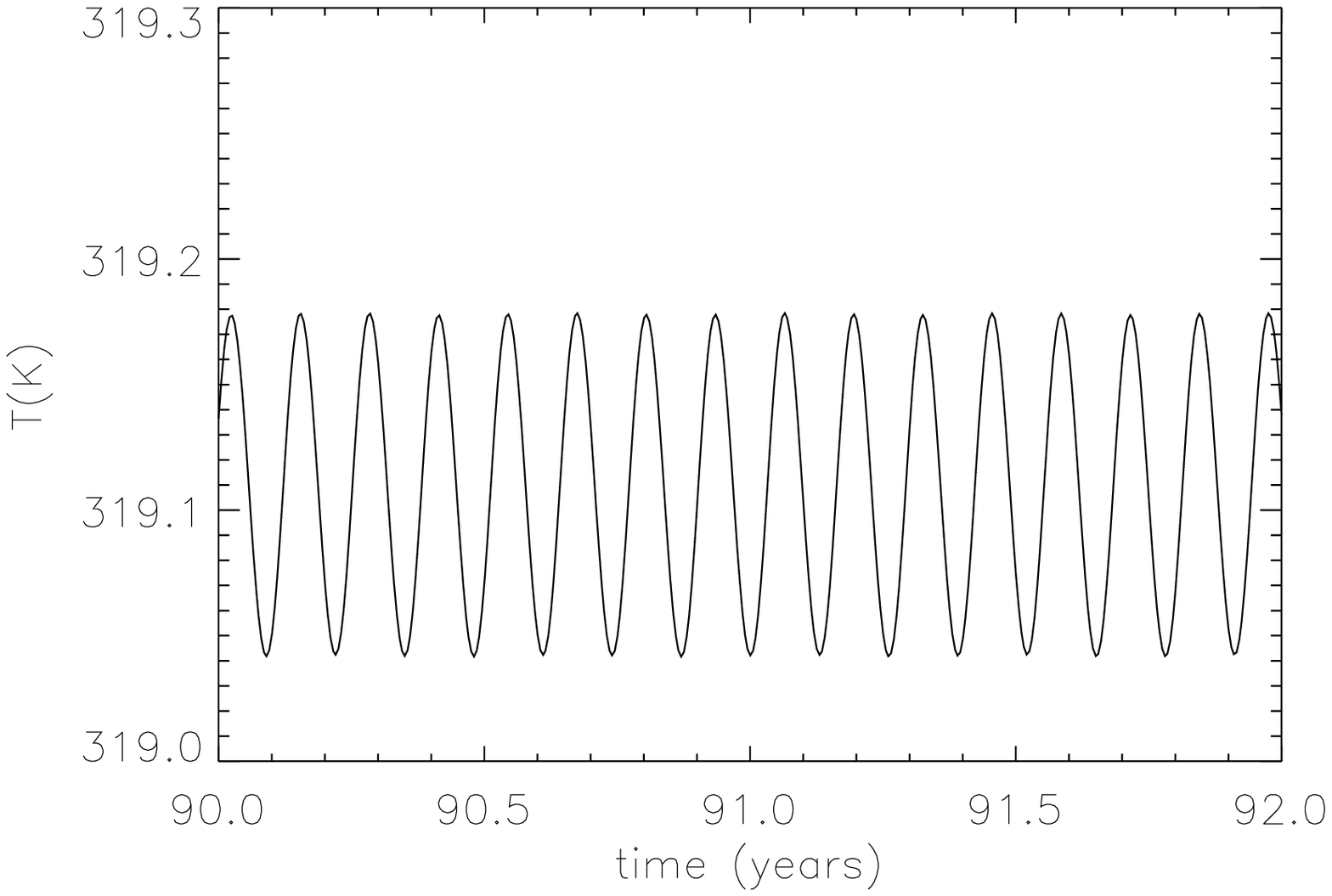} &
\includegraphics[scale=0.4]{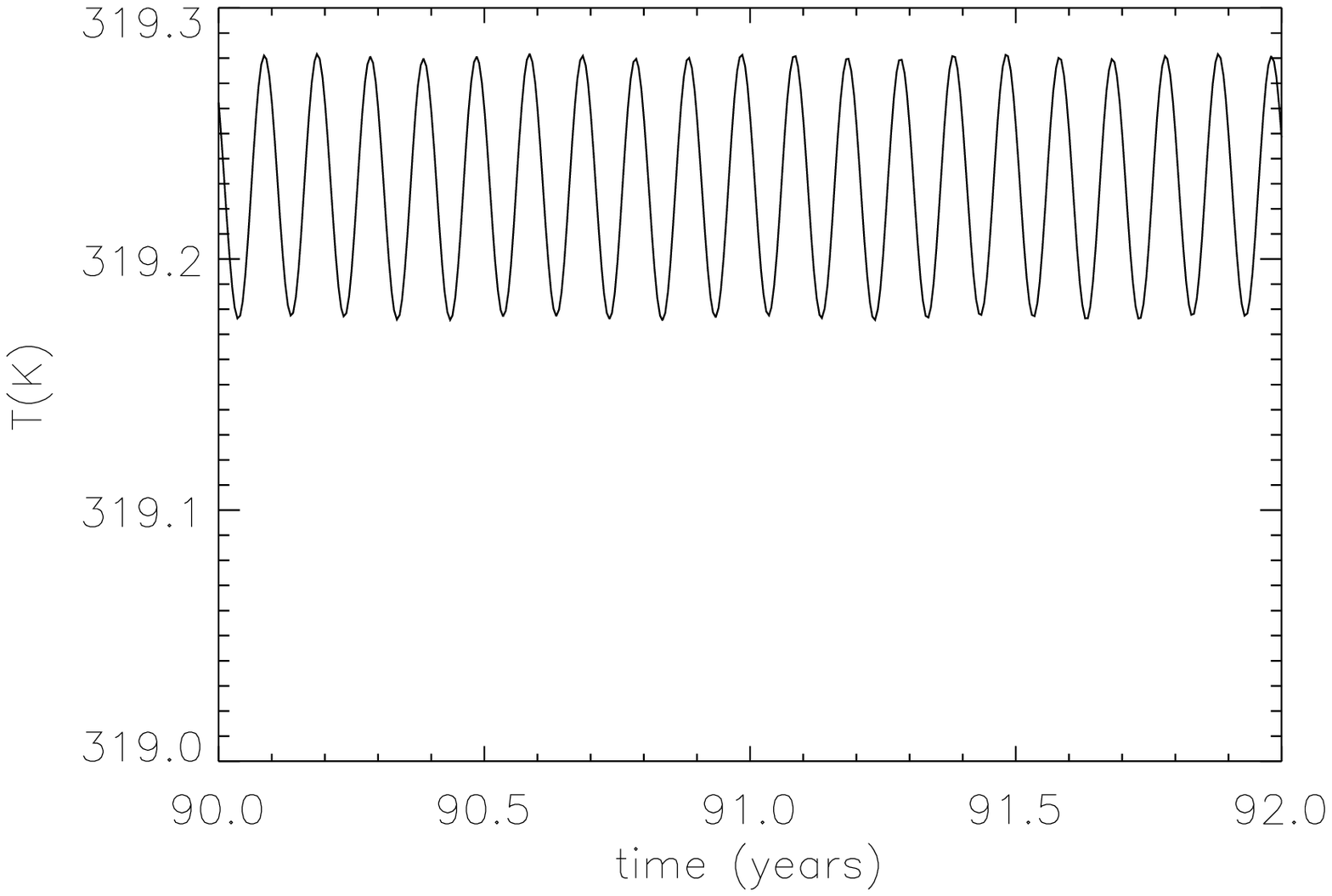} \\
\includegraphics[scale = 0.4]{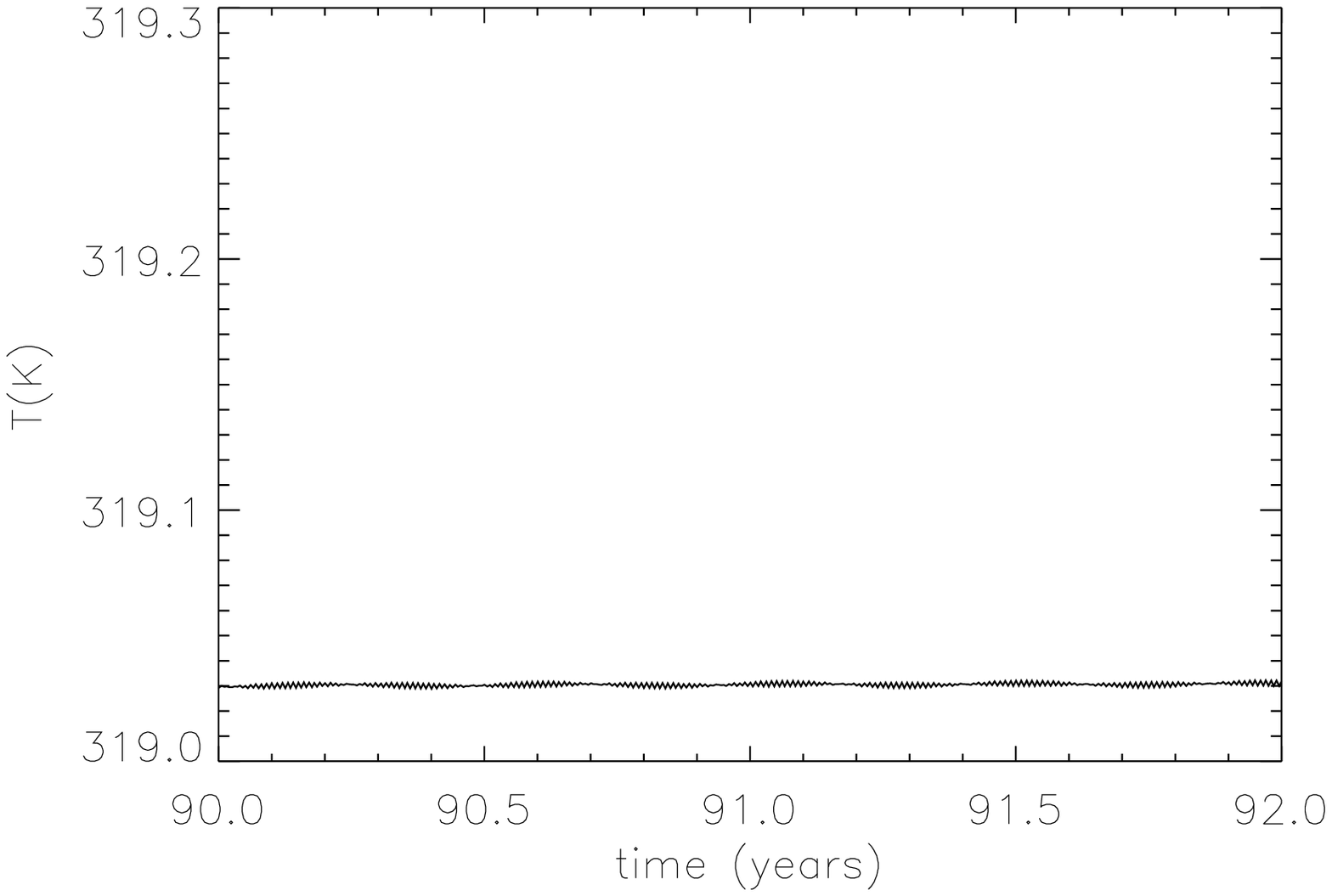} &
\includegraphics[scale=0.4]{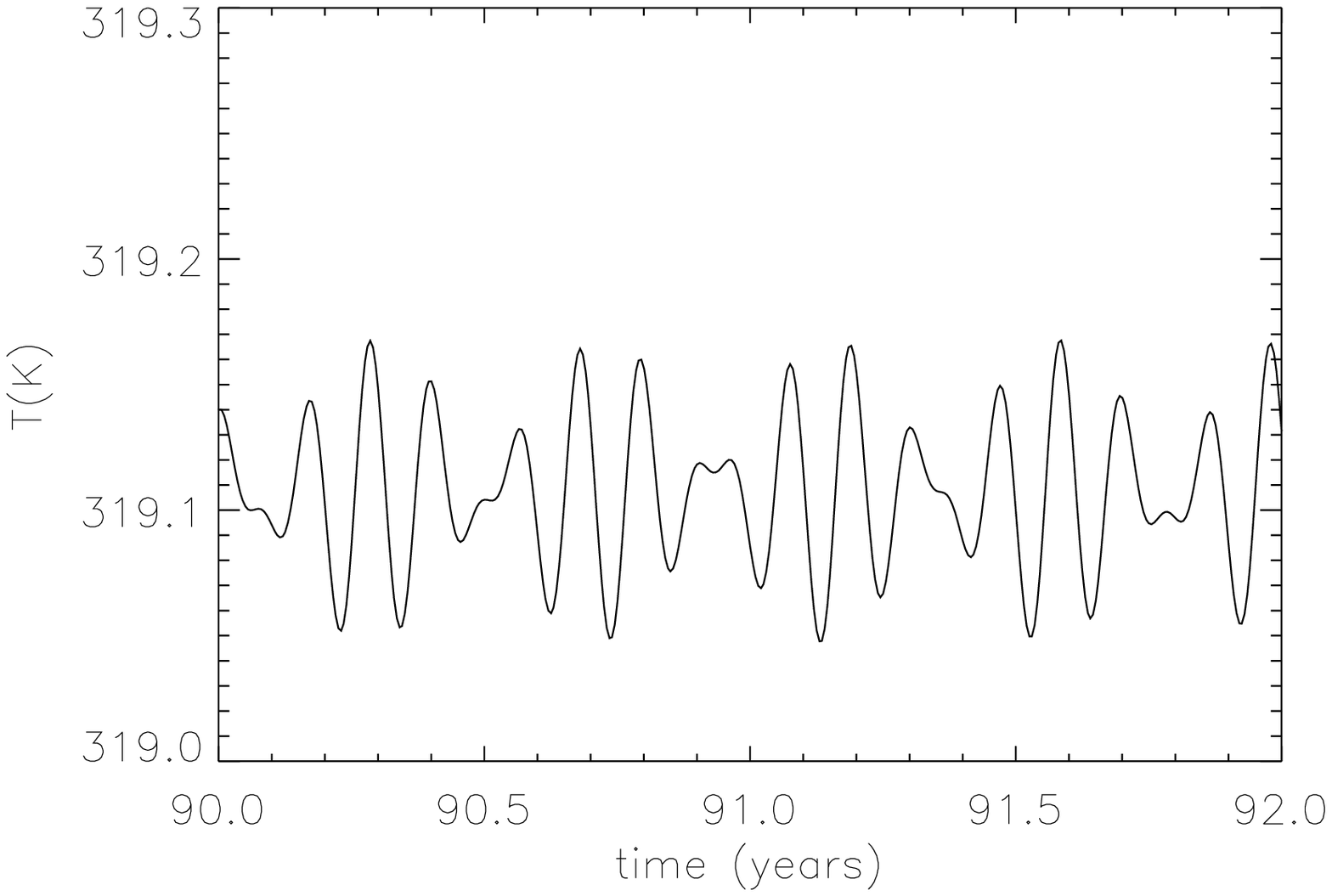} 
\end{array}$
\caption{The mean temperature of the four test cases, over 2 years, approximately 18 moon orbital periods (560 for case 3), and approximately 2.35 planetary orbital periods).  The top left panel shows case 1, the top right panel shows case 2, the bottom left panel shows case 3, and the bottom right panel shows case 4.  \label{fig:casesmeanT}} 
\end{center}
\end{figure*}

Why should we see such a large change by changing something as simple as the direction of orbit, and yet see little change by reducing the moon's orbital distance by a factor of 5?  The answer lies in considering the mean temperature more carefully (Figure \ref{fig:casesmeanT}).  The mean temperature fluctuates by $<0.1$ K, in all cases, but it is the nature of the fluctuations that are telling.  The retrograde orbit is consistently hotter than the prograde orbit, and the frequency of temperature fluctuations is also higher.  Case 4, where the exomoon orbits perpendicular to the plane, exhibits a superposition of frequencies, and case 3, with low $a_m$, exhibits essentially no variation at all.

We can find an explanation for these phenomena if we consider the orbital dynamics.  Let us consider first the difference between the prograde and retrograde cases.  

\begin{figure}
\begin{center}
\includegraphics[scale=0.3]{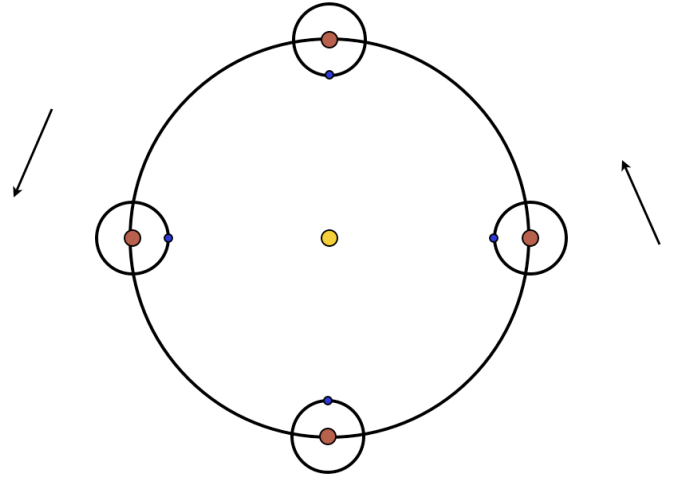}
\caption{The precession of the moon's longitude of periastron.  The sun and planet are marked in yellow and red respectively, with the moon's longitude of periastron indicated by the blue dot. \label{fig:precess}}
\end{center}
\end{figure}

The planet's orbit is circular, and as such has no apastron or periastron.  The moon's orbit is also circular relative to the planet, and hence has no apojove or perijove.  But, relative to the star it undergoes epicyclic motion, and hence the moon has well-defined periastron and apastron.  Figure \ref{fig:precess} shows the location of the moon's periastron over the course of the planet's orbit.

The longitude of moon periastron, $\phi_{\rm per}$, precesses with a period equal to the planet's orbital period, $P_{\rm  pl}$.  We define the x-axis as corresponding to $\phi=0$, and therefore initially, $\phi_{\rm per}= \pi $ radians. So,

\begin{equation} 
\phi_{\rm per}(t) = \pi + \dot{\phi}_{\rm pl} t \, \mathrm{mod}\, 2\pi .
\end{equation}

\noindent If the planet were static ($\dot{\phi}_{\rm pl} =0$) then the longitude of periastron would not precess, and the time taken for the moon to move from apastron to periastron would be 

\begin{equation} 
\tau_0 = \frac{\pi}{\dot{\phi}_{\rm m}} ,
\end{equation}

\noindent where $\dot{\phi}_{\rm m} = 2\pi /P_m = n_m$, and $P_m$ is the moon's orbital period. However, $\dot{\phi}_{\rm pl} \neq 0$, and therefore the longitude of periastron moves during this interval.  If the moon's orbit is prograde, the angular distance between apastron and periastron increases, and the timescale becomes

\begin{equation} 
\tau_P = \frac{\pi + \dot{\phi}_{\rm pl}\tau_P}{\dot{\phi}_{\rm m}} .
\end{equation}

\noindent If the orbit is retrograde, the angular distance decreases, and

\begin{equation} 
\tau_R = \frac{\pi - \dot{\phi}_{\rm pl}\tau_R}{\dot{\phi}_{\rm m}} .
\end{equation}

\noindent Rearranging the above equations gives

\begin{equation} 
\tau_P = \frac{\pi}{\dot{\phi}_{\rm m} - \dot{\phi}_{\rm pl}} 
\end{equation}

\noindent and

\begin{equation} 
\tau_R = \frac{\pi}{\dot{\phi}_{\rm m} + \dot{\phi}_{\rm pl}} 
\end{equation}

\noindent respectively.  Therefore, the ratio of one to the other is 

\begin{equation}
\frac{\tau_R}{\tau_P} = \frac{\dot{\phi}_{\rm m} - \dot{\phi}_{\rm pl}} {\dot{\phi}_{\rm m} + \dot{\phi}_{\rm pl}}
\end{equation}



\noindent The orbital periods of both bodies are related by (Appendix B of \citealt{Kipping2009}):

\begin{equation}
P_{\rm m} = P_{\rm pl} \sqrt{\frac{\chi^3}{3}},
\end{equation}

\noindent where $\chi$ is $a_m$ normalised by the moon's Hill Radius.   This gives the more digestible

\begin{equation}
\frac{\tau_R}{\tau_P} = \frac{\sqrt{3} - \chi^{3/2}}{\sqrt{3} + \chi^{3/2}}
\end{equation}

\noindent For the object masses and semi-major axes used here, this corresponds to a ratio of 0.758.  Therefore, the forcing timescale of retrograde moons is faster than prograde exomoons, and hence we might expect to see more variations in temperature for these orbits.  Comparing the period of oscillations for cases 1 and 2, we find the ratio is indeed approximately 0.758.

We should also note the amplitude of the oscillations, with the prograde displaying an amplitude of around 0.14 K, compared to the retrograde amplitude of 0.1K.  This is again sensible considering the orbital dynamics: $\tau_P>\tau_R$,  the prograde moon will spend more time at larger and smaller distances from the star, allowing the mean temperature to
increase and decrease appropriately.  The retrograde moon will encounter apastron and periastron more frequently, but spend less time there.  This will not allow the mean temperature to rise as much, but will facilitate some latitudes to increase and decrease their temperature more readily (which gives an explanation for the high oscillations seen in the maximum and minimum temperatures in Figure \ref{fig:casescompareT}). 

Case 3 has a significantly lower orbital period (approximately 3.6 days), compared to the 41 day orbit of the other moons.  As the diurnal and orbital periods are extremely close, and the periastron and apastron distances are extremely similar, the effect of epicyclic motion is extremely weak, and the climate becomes difficult to distinguish from that of an Earth-like planet orbiting with parameters $(a_p,e_p)$.

\begin{figure}
\begin{center}
\includegraphics[scale=0.3]{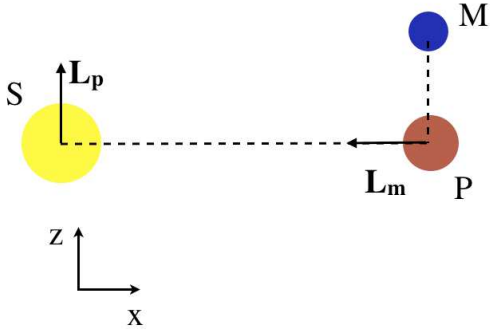}
\caption{Arrangement of $L_p$ and $L_m$, the planet and moon orbital angular momentum vectors respectively, in test case 3, a polar orbit.  The figure shows the arrangement at time $t=0$ \label{fig:polarorbit}}
\end{center}
\end{figure}

The counter-intuitive behaviour of Case 4 becomes clear when the alignment of the two orbital angular momentum vectors, $\mathbf{L_p}$ and $\mathbf{L_m}$ are considered.  $\mathbf{L_p}$ is aligned with the z-axis, and $\mathbf{L_m}$ is aligned with the x-axis in the negative direction (see Figure \ref{fig:polarorbit}).  At $t=0$, the vector product $\mathbf{L_p} \times \mathbf{L_m} = \mathbf{L'}$ is aligned with the negative y direction.  As there are no external torques acting on the system, $\mathbf{L'}$ maintains this alignment throughout the orbit, and as such the moon's periastron and apastron distances become a function of the orbital longitude of the planet.

At $t=0$, the periastron and apastron distances are equal, and the distance between the star and the moon $r_{\rm *m}$ will not change during its orbit.  After 0.25 planetary orbital periods have elapsed, the periastron and apastron distances are now equal to that in the other 3 cases.  Similarly, at $t=0.5$ planetary orbital periods the periastron and apastron distances are the same, and at $t=0.75$ planetary orbital periods they reach their maximum values once more.  This provides the amplitude modulation seen in the bottom right panel of Figure \ref{fig:casesmeanT}.  Note the frequency of the oscillation is the same as that of case 1, as they are both prograde orbits. 

\subsection{Parameter study}

\noindent The previous section has shown us that while the global properties of the exomoon are relatively insensitive to the orbital architecture, the detailed behaviour of the exomoon's climate depends on the orbital direction and period.  We shall now look at how the global properties depend on the parameters more easily determinable by exoplanet and exomoon searches: the semi major axes of both objects relative to their host, and their eccentricities.  To do this, we carry out over 3500 separate climate simulations, and classify each one as hot, cold, habitable or transient according to the classification system outlined previously.

\subsubsection{A Control - The Circumstellar Planetary Habitable Zone}

\begin{figure*}
\begin{center}$
\begin{array}{cc}
\includegraphics[scale = 0.4]{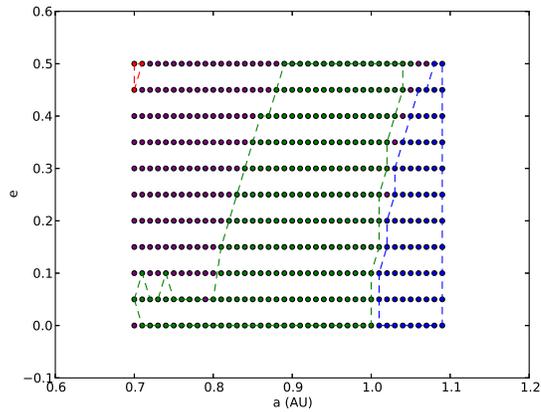} 
\end{array}$
\caption{The habitable zone for an Earth-like planet around a Sun-like star, as calculated from a LEBM using the classification system outlined in this paper, for comparison with the exomoon habitable zones displayed in later figures. Each point represents a simulation run with these parameters, and the colour of the point indicates its outcome.  Red points produce hot planets with no habitable surface; blue points produce cold planets with no habitable surface; green points represent warm planets with at least ten percent of the surface habitable and low seasonal fluctuations; purple points represent warm moons with high seasonal fluctuations. \label{fig:earth_ae_space}}
\end{center}
\end{figure*}

To provide grounds for comparison, we ran a separate series of 440 simulations with an Earth-like planet orbiting a Sun-like star to map out the traditional circumstellar HZ according to our classification system.  The results of this are displayed in Figure \ref{fig:earth_ae_space}.  Green points represent simulations where the planet is classified as habitable; purple points where the planet is classified as a transient; and the red and blue points indicate planets that are too hot and too cold respectively.

The model does not contain a carbonate-silicate cycle, unlike e.g. \citealt{Williams1997a}, which modified the atmospheric $CO_2$ pressure in accordance with temperature dependent weathering rates.  Lower temperatures produce lower weathering rates, and as a result the atmospheric $CO_2$ (produced by volcanic outgassing) cannot be sequestered.  Therefore, cooler planets can be expected to have higher atmospheric concentrations of $CO_2$, boosting the greenhouse effect and moving the outer edge of the HZ to higher semi-major axes than we see in Figure \ref{fig:earth_ae_space}  (cf \citealt{Kasting_et_al_93}).

The extension of the HZ at low $e_p$ to semi-major axes as low as 0.7 AU is a reflection of our (fairly lenient) criterion for habitability - namely, that 10\% of the planet's surface remains habitable over a ten year period, with a standard deviation less than 10\% of the mean habitable area.  As $e_p$ is increased, $\sigma_\xi$ increases quickly, producing planets which are habitable on a seasonal basis only.  If we are to compare to traditional habitability studies, then we should infer that the inner edge of the HZ is at approximately 0.8 AU for $e_p=0$.  Equally, many of the transient classifications in this study would normally have been considered to be uninhabitable (as many of these simulations undergo seasonal periods when the habitable surface fraction is close to zero, but this is not sufficient to label them as hot or cold planets).  We should bear this in mind as we consider the habitable zone for exomoons in the following sections.

\subsection{Zero Moon Orbital Eccentricity}

\begin{figure*}
\begin{center}$
\begin{array}{cc}
\includegraphics[scale = 0.4]{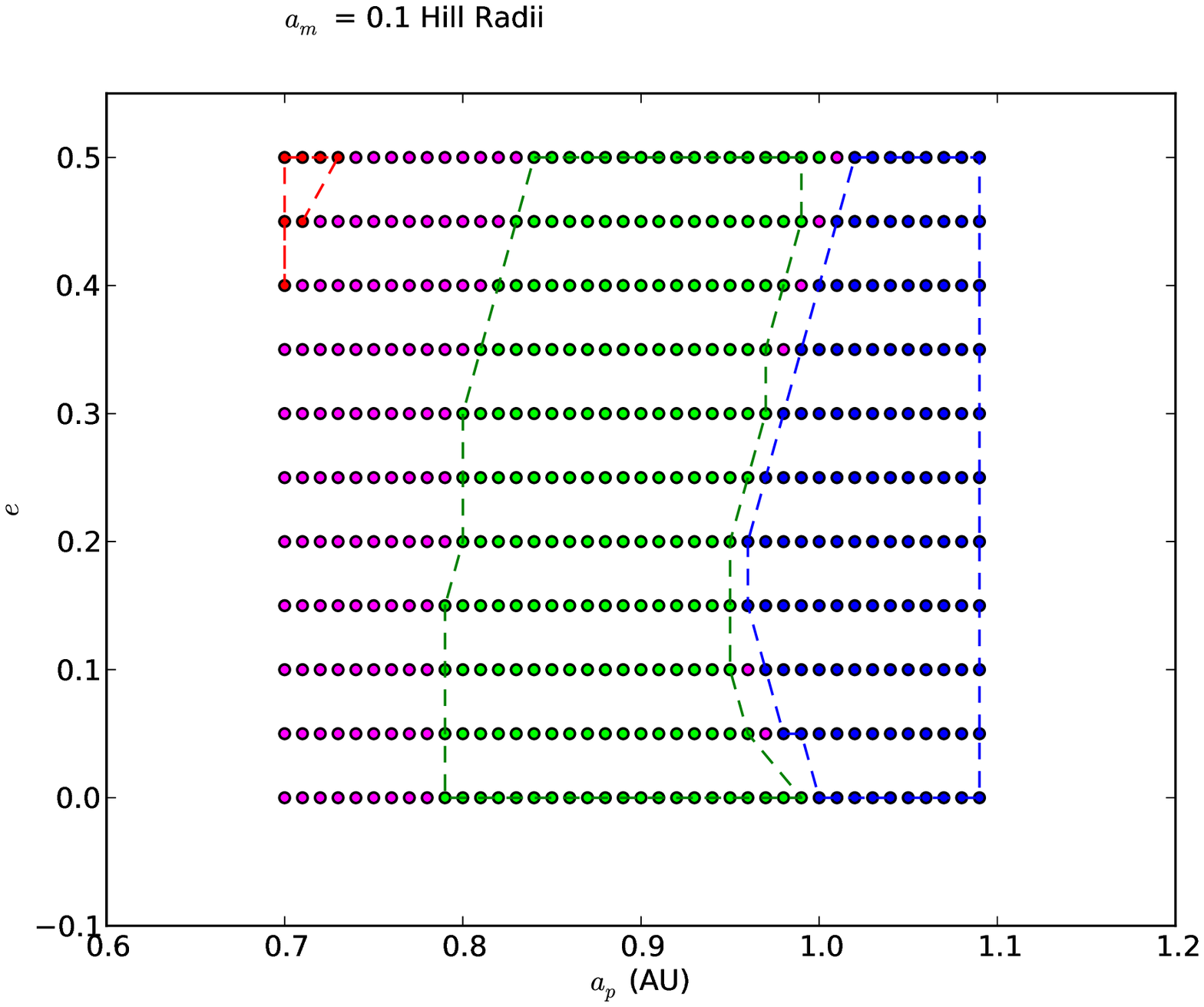} &
\includegraphics[scale=0.4]{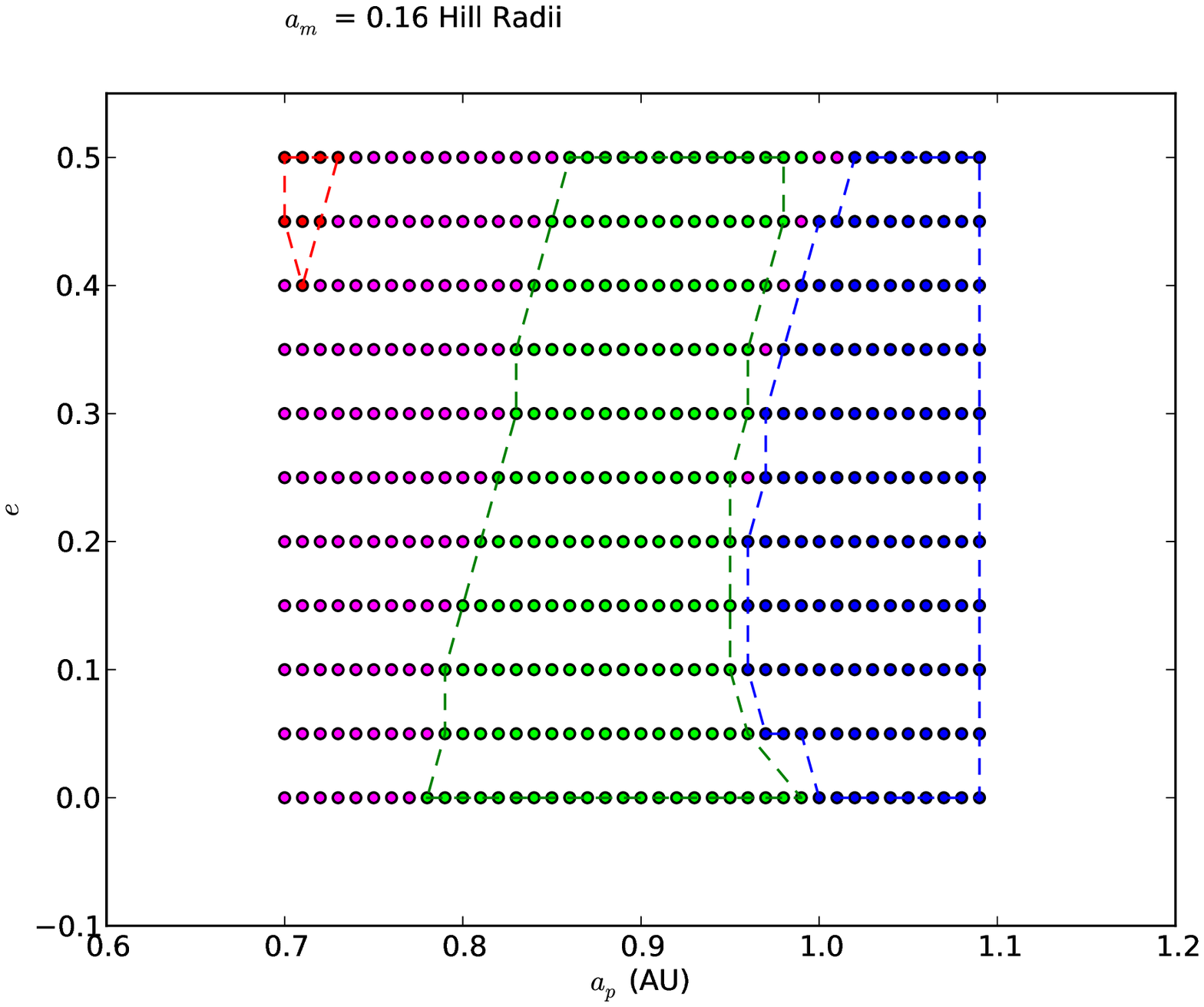} \\
\includegraphics[scale = 0.4]{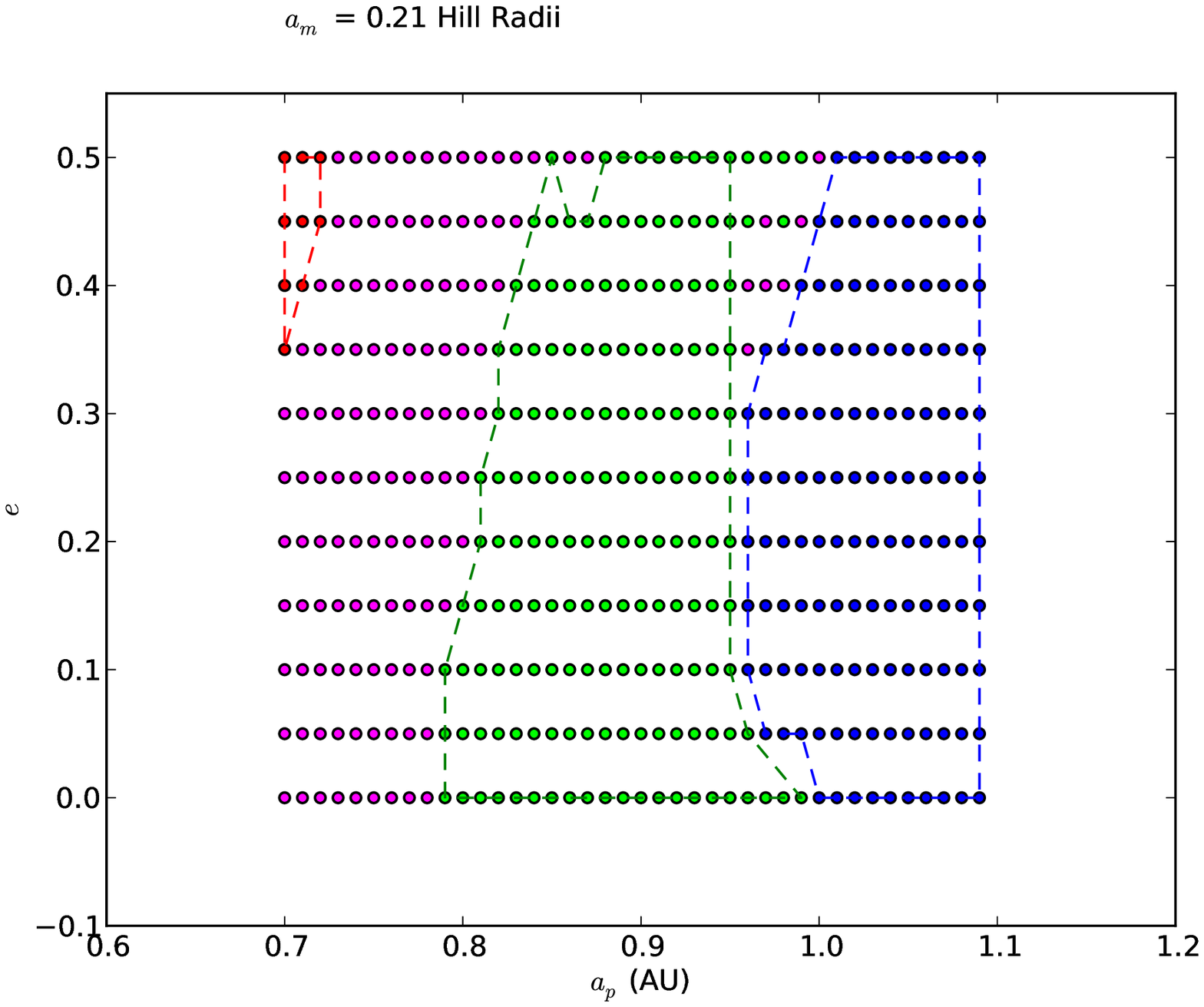} &
\includegraphics[scale=0.4]{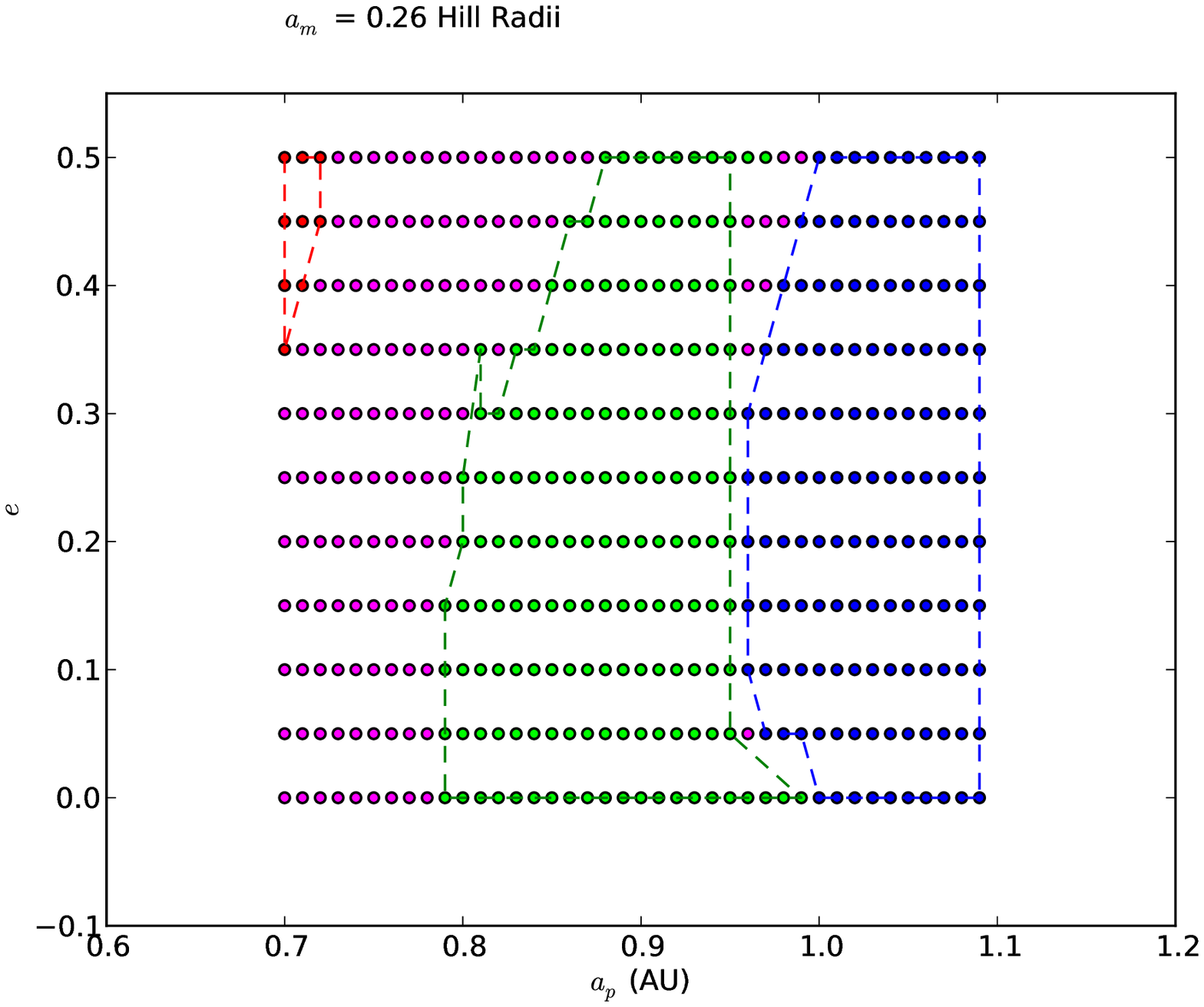} 
\end{array}$
\caption{The exomoon habitable zone, as a function of host planet semi-major axis $a_p$ and eccentricity $e_p$, for moons with zero orbital eccentricity.  Each point represents a simulation run with these parameters, and the colour of the point indicates its outcome.  Red points produce hot moons with no habitable surface; blue points produce cold moons with no habitable surface; green points represent warm moons with at least ten percent of the surface habitable,  and low seasonal fluctuations; purple points represent warm moons with high seasonal fluctuations.  In each plot, the moon's orbital semi-major axis relative to the host planet, $a_m$ is fixed at a value between 0.1 and 0.3 Hill Radii. \label{fig:apep_em0}}
\end{center}
\end{figure*}

\noindent We first consider the habitable zone manifold in the case where $e_m=0$, (and consequently the tidal heating rate is zero) and we consider several different values of $a_m$.  This allows us to construct $a_p-e_p$ slices of the HZ manifold to compare with the typical planet HZ shown in the previous section.  Figure \ref{fig:apep_em0} shows four $a_p-e_p$ slices, for four different values of $a_m$.  In each simulation, we fix $a_m$ relative to the Hill Radius at the given value of $a_p$, so the absolute value of $a_m$ will increase with increasing $a_p$.  

Comparing the top left panel of Figure \ref{fig:apep_em0} with Figure \ref{fig:earth_ae_space}, we can immediately see that the habitable zone exists in general at lower semi-major axis.  This is especially true at high $e_p$: the inner edge of the HZ at $e=0.5$ is now at 0.84 AU, as compared to 0.89 AU in the planetary case.  As $a_m$ is increased, the high $e_p$ component of the inner HZ boundary shifts to higher and higher $a_p$.  We therefore surmise that the effect of frequent eclipses allows the moon to remain cooler, and damps the fluctuations in temperature experienced as the planet eclipses the star.

In all cases, the low $a_p$, low $e_p$ habitable tail observed in the planetary case disappears in the exomoon case.  While frequent eclipses can cool the climate and damp oscillations, the habitable surface fraction of these moons is generally low (typically close to the minimum 0.1).  Therefore, a relatively small increase in the value of $\sigma_\xi$ due to eclipses is large enough to ensure $\sigma_\xi > 0.1 \bar{\xi}$, and as a result be classified as transient moons rather than habitable moons.

Comparing the outer edge of the HZ in the planet and moon cases, we can see it takes on a significantly different shape.  While the planet outer HZ moves steadily outward as $e_p$ is increased, the moon outer HZ changes direction as $e_p$ is increased.  Initially habitable at 0.99 AU for $e_p=0$, the outer boundary moves inward as $e_p$ increases, finding its minimum value of $\sim 0.95$ AU at $e_p=0.2$, and then moves outward again.  This appears to be due to a combination of planet and moon motion relative to the star.  The moon's epicyclic motion at $e_p=0$ changes its distance from the star by only a small amount, and hence the climate remains stable enough to be classified as habitable.  As $e_p$ is increased, the epicyclic motion of the moon is modulated by the motion of the planet as it moves between its own periastron and apastron.  This extra modulation is sufficient to trigger a positive feedback in albedo, and a subsequent snowball effect.  As $e_p$ increases to large values, the strong heating the moon experiences as the planet passes through its periastron is sufficient to keep it warm throughout the rest of the orbit, and the outer HZ boundary remains at $a_p \sim 0.99$ AU.  As the eclipse duration increases with increasing $a_m$, the outer HZ boundary is pushed inwards for larger $e_p$ as $a_m$ is increased.

\subsection{High Moon Orbital Eccentricity}

\begin{figure*}
\begin{center}$
\begin{array}{cc}
\includegraphics[scale = 0.4]{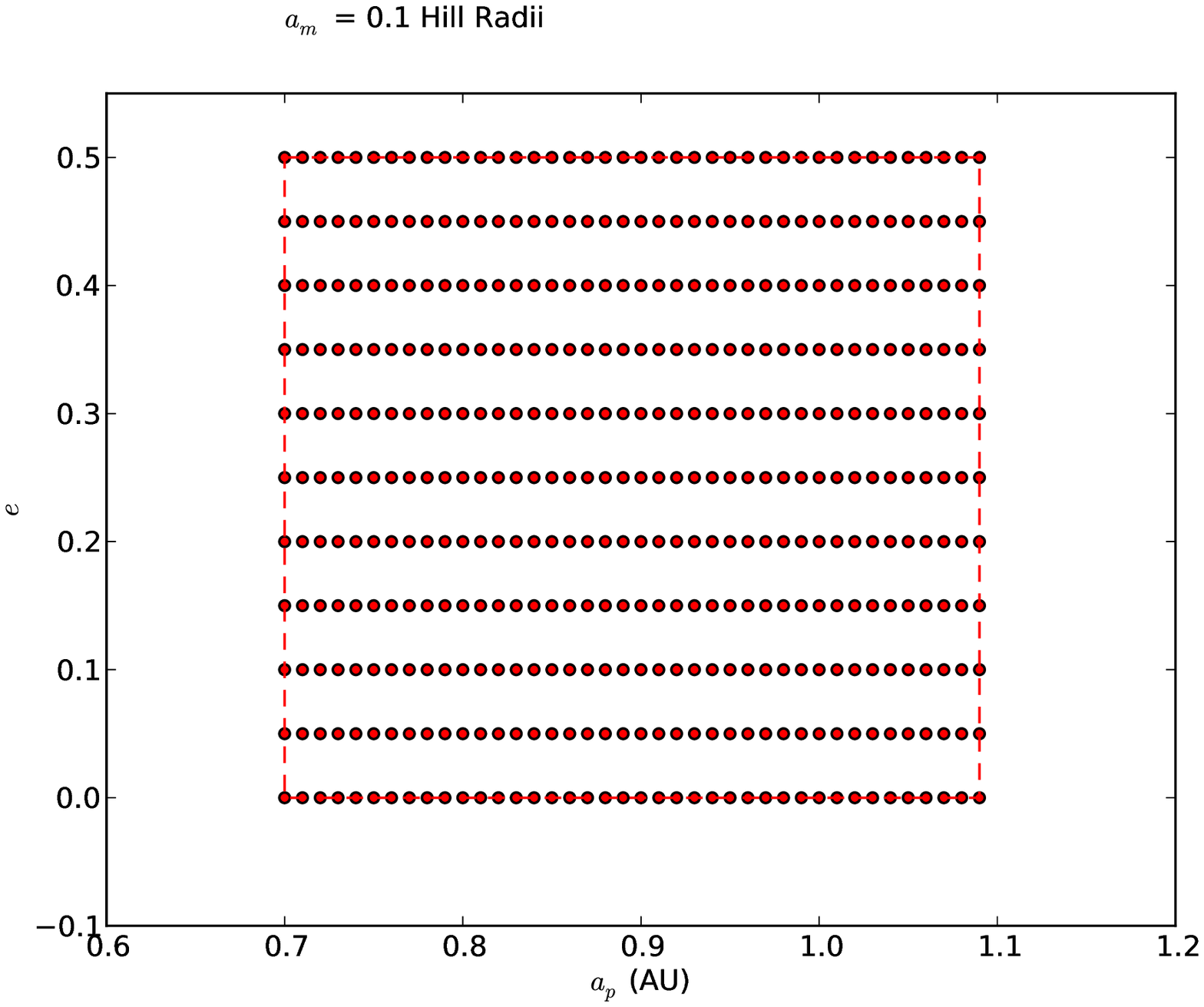} &
\includegraphics[scale=0.4]{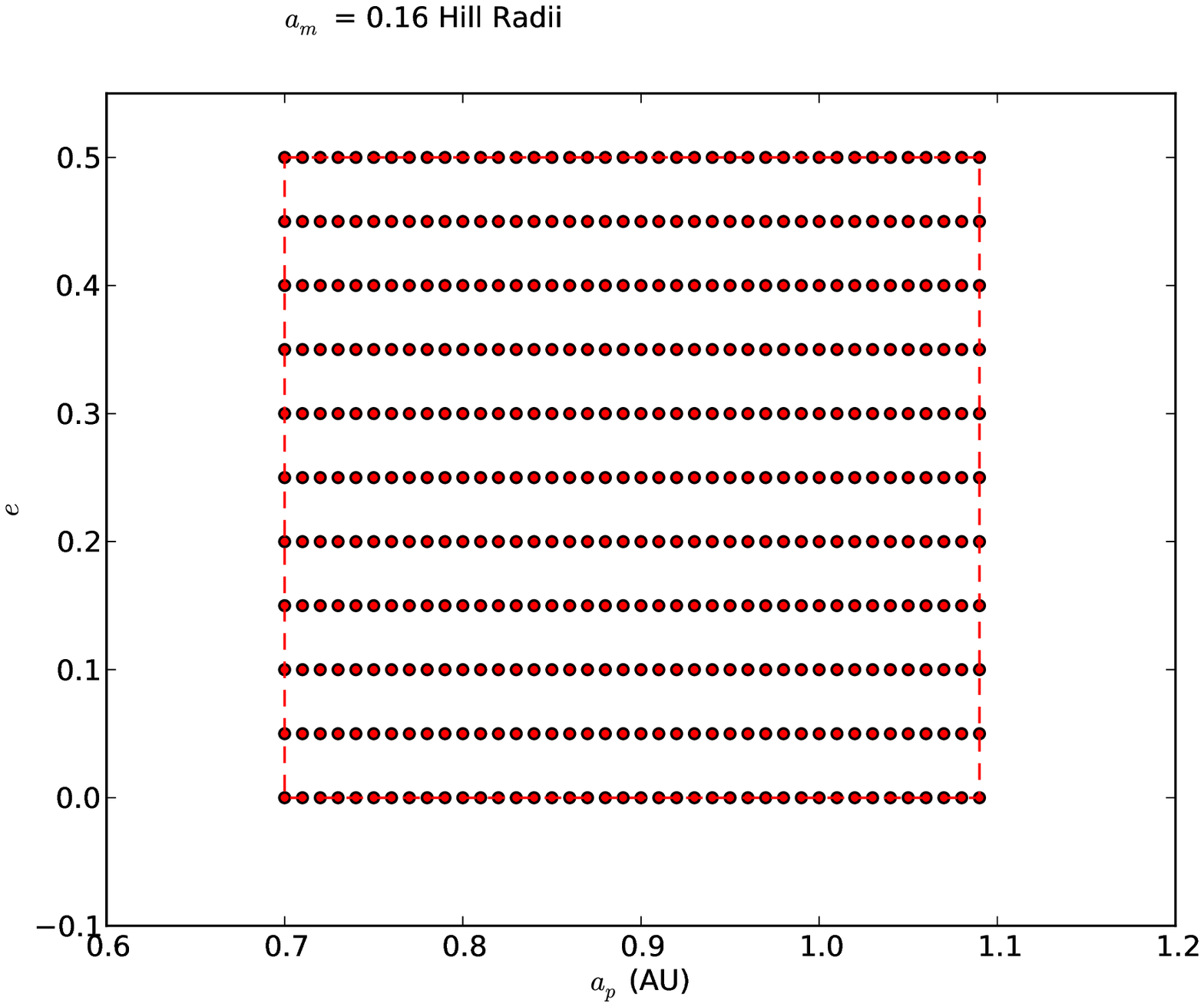} \\
\includegraphics[scale = 0.4]{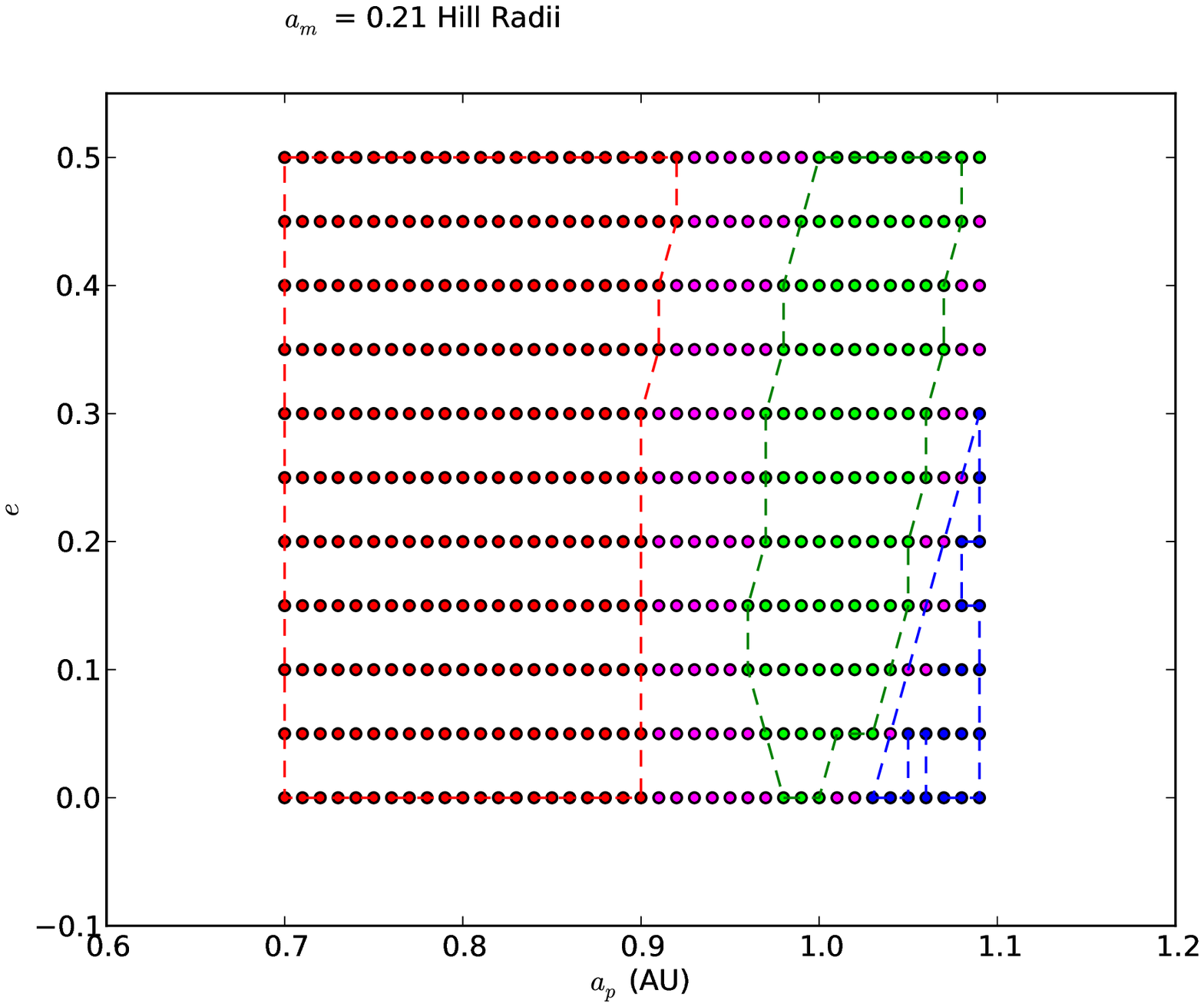} &
\includegraphics[scale=0.4]{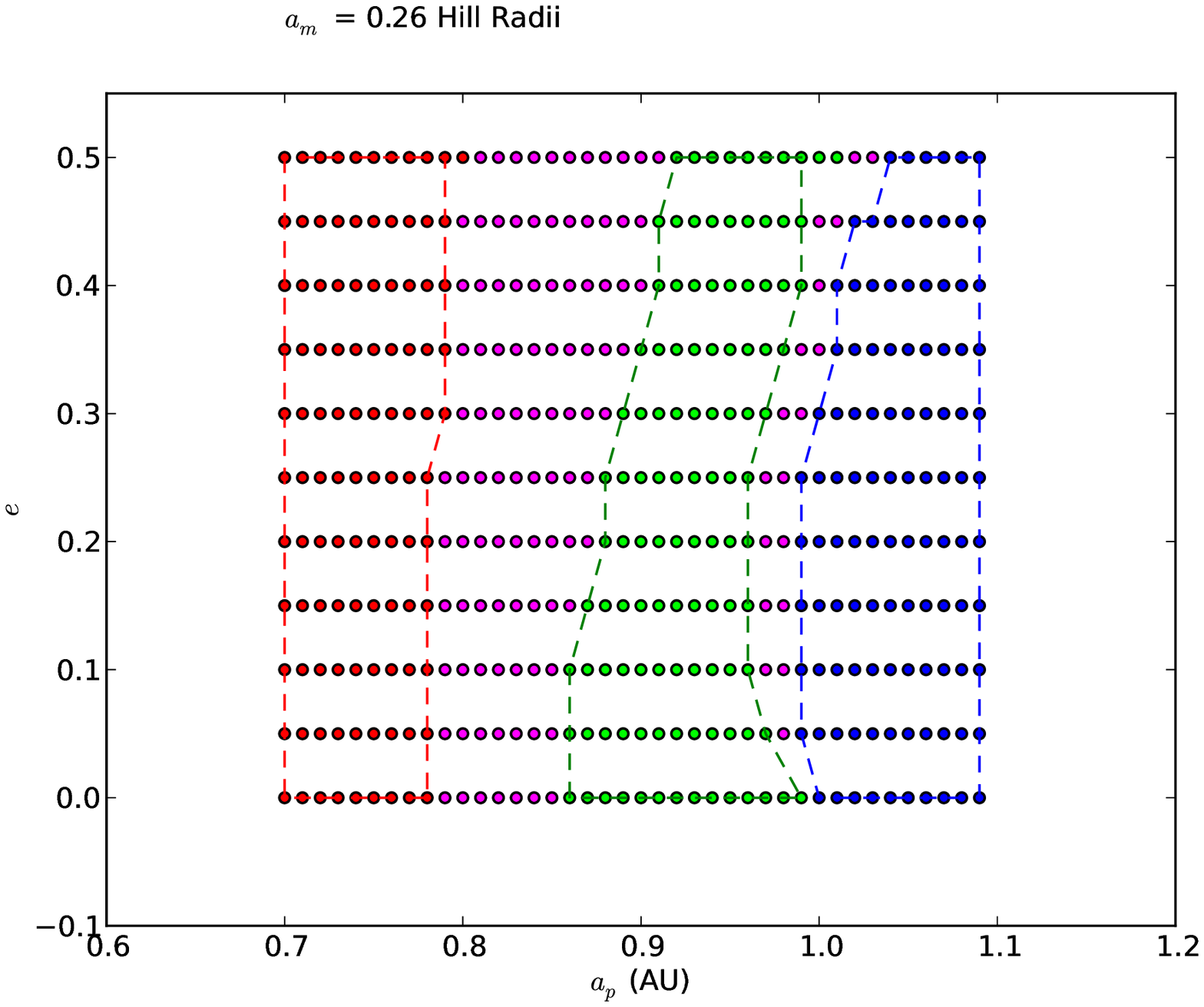} 
\end{array}$
\caption{The exomoon habitable zone, as a function of host planet semi-major axis $a_p$ and eccentricity $e_p$, for moons with eccentricity $e_m=0.1$.  Each point represents a simulation run with these parameters, and the colour of the point indicates its outcome.  Red points produce hot moons with no habitable surface; blue points produce cold moons with no habitable surface; green points represent warm moons with at least ten percent of the surface habitable, and low seasonal fluctuations; purple points represent warm moons with high seasonal fluctuations.  In each plot, the moon's orbital semi-major axis relative to the host planet, $a_m$ is fixed at a value between 0.1 and 0.3 Hill Radii.\label{fig:apep_em01}}
\end{center}
\end{figure*}

\noindent We now repeat the above analysis, but increase the moon's eccentricity to $e_p=0.1$.  This is quite a large value to assume for moon eccentricity - without eccentricity pumping, it is unlikely that moons will possess eccentricities as large as this after tidal evolution with their host planet - we present these results as an extreme case.  Intermediate eccentricities will produce results somewhere between the results of this section and the results of the previous section.

Indeed, Figure \ref{fig:apep_em01} illustrates how extreme this case is by the effect on the climate of the simulations.  As tidal heating increases as $a_m$ is decreased, it is clear that tidal heating dominates the properties of highly eccentric moons for $a_m < 0.2$ Hill Radii (top panels), and all simulations are classified as hot.

As $a_m$ is increased, a habitable zone appears.  This zone is much narrower than seen previously, and is restricted to large values of $a_p$, as tidal heating allows for warmer solutions when insolation is weak.  In the case where $a_m = 0.21 $ Hill Radii (bottom left panel of Figure \ref{fig:apep_em01}), the HZ exists at $a_p > 0.95$ AU, and potentially extends beyond 1.1 AU at high $e_p$.  The HZ is typically around 0.1 AU in width, although at low $e_p$ it is as narrow as 0.03 AU.  

When $a_m$ is increased to 0.26 Hill Radii, the HZ begins to more closely resemble the HZ seen for $e_m=0$ moons.  While still around 0.1 AU in width, it now displays the knee in the outer HZ boundary seen previously, and it is centred on similar values of $a_p$ as before.  At this value of $a_m$, the tidal heating has become small compared to the traditional heating from insolation.  This being the case, tidal heating still makes its presence felt at low $a_p$, where simulations that would have been habitable for $e_p=0$ are now seasonally habitable due to the extra forcing from tidal heating producing stronger climate oscillations.

\section{Discussion}\label{sec:discussion}

\subsection{Limitations of the Model}

As with all studies using 1D LEBMs, the nature of the assumptions made limit the applicability of the results obtained in this analysis.  The models are insensitive to the longer term climate variations produced by oceanic circulation, as well as the compensatory effects of the carbonate-silicate cycle which can halt the albedo-driven snowball effect.  These effects, with timescales ranging between a few thousand to a million years \citep{Williams1997a} are much longer than the orbital periods of both the moon and the host planet, and as such are of less consequence (in the short term) than the dynamical forcing due to eclipses and tidal heating.

Assuming a value for the diffusion constant, $D$ limits the application of LEBMs to exomoons in a different sense.  $D$ is calibrated such that a fiducial Earth-Sun climate model produces the correct latitudinal temperature distribution as obtained from real data \citep{Spiegel_et_al_08}.  Therefore, $D$ contains a great deal of hidden information about the atmospheric and thermodynamic properties of the Earth.  Modifying these properties, e.g. the atmospheric pressure, has significant consequences for the habitability of the object \citep{Williams1997a, Vladilo2013} and it is less clear how $D$ ought to be modified for frozen moons such as Europa, or moons with fundamentally different chemistry such as Titan, without more data on their own latitudinal temperature distributions.  Even with this data, a full understanding of how tidal heating affects the temperature balance at any latitude is essential for the correct calibration of $D$.  Equally, a more accurate implementation of tidal heating in the LEBM is vital for future studies of exomoons.  

Tidal forces are expected to affect land and ocean according to their elastic rigidities.  While we have assumed a fixed ocean fraction of 0.7 throughout this analysis, the LEBM does not contain information as to how these oceans are distributed across the surface.  Instead, the ocean and land components are assumed to be uniformly mixed over all latitudes.  Future calculations of tidal heating will require the LEBM to be more specific about the distribution of landmasses.  Ocean fraction may therefore become an important parameter in future studies of the exomoon HZ manifold (which should also investigate the effects of changing other parameters such as obliquity).  Besides its effect on tidal heating, the ocean fraction also affects the thermal relaxation timescale of the moon - reducing the ocean fraction leaves the moon more susceptible to larger climate oscillations \citep{Spiegel_et_al_08,Abe2011,Forgan2012}. 

Tied to this problem of limited surface data is the limited dimension of the model itself - ignoring longitudinal information requires us to assume the moon is rapidly rotating relative to the insolation source.  This will remain true for exomoons which are tidally locked \emph{to the planet} - as they will still rotate with respect to the primary insolation source, the star - but the more complicated nature of the lunar day compared to a planetary day may mean that the rapid rotation assumption is satisfied more weakly at some points in the orbit compared to others.  

We have also ignored one important heat source - illumination of the moon by the planet.  \citet{Heller2013} show calculations for illumination from a tidally locked planet, incorporating the planet's thermal radiation and reflected starlight, with reflected starlight typically dominating unless the planet's albedo is very low.  Adding the reflected starlight alone can increase the outer HZ boundary by a few percent - if this was incorporated into the LEBM, it could act to prevent the "knee" in the outer boundary at moderate values of $e_p$.  

Finally, we assume orbital parameters for all objects involved, and we do not evolve these parameters under the presence of any gravitational field.  Incorporating the LEBM component into a more involved computation of the tidal evolution of the star-planet-moon system (e.g. \citealt{Laskar1993, Sasaki2012}) would allow an investigation of a unique set of Milankovitch-esque cycles produced by the moon's orbital evolution (see e.g. \citealt{Spiegel2010}), with an accurate computation of the moon's obliquity evolution and tidal heating as a bonus.

\subsection{Implications for Observations}

Our simulations suggest that determining the orbital eccentricity of any exomoons detected in or near the conventional HZ will be crucial in assessing whether the moon itself is also habitable, much more so in fact than determining the host planet's eccentricity. Exomoons with even a moderate amount of eccentricity of $\sim$0.1 lead to a more limited habitable-zone with respect to the other system parameters as a result of tidal heating dominating. One may consider that the HZ extends outwards in $a_P$ for such scenarios but longer period planets are considerably rarer in the known transiting exoplanet catalogue, due to both geometric bias and lower composite signal-to-noise.

The orbital eccentricity of a moon will affect the position and duration of moon transits in the light curve \citep{Kipping2011} and is, in principle, an observable quantity. It is therefore quite reasonable that constraints on exomoon eccentricity could be determined in the case of a confirmed detection. It is also worth noting that exomoons are expected to rapidly tidally circularize even if starting with a large initial eccentricity due to a capture, for example \citet{porter2011, Williams2013}.

To maintain a large exomoon eccentricity over Gyr would likely require forcing due to resonant massive satellites, but the occurence of multiple captures of Earth-mass moons around a single planet is surely less than that of single captures of Earth-mass moons.

We also find that the sense of orbital motion affects the climate of moons and in principle this may also be determined observationally \citep{Kipping2009a}. However, in general such a determination
is highly challenging with \citet{Kipping2012} finding even optimstic synthetic data is marginal for uniquely determining this to high-confidence using \emph{Kepler}. Of course, once a confirmed signal has been found, follow-up with larger instrumentation such as JWST would greatly increase
our ability to make this assesment.

Finally, in this work we have established an LEBM capable of modelling hundreds of realizations of exomoon configurations. Due to the likely low signal-to-noise of any future exomoon detection one should expect fuzzy and complex parameter posteriors, as demonstrated in synthetic
tests by \citet{Kipping2012}. We therefore suggest coupling an LEBM, such as that presented here, directly with representative realizations drawn from the joint posterior distribution of the relevant parameters.  This would enable us to compute a marginalised distribution for the habitability of an exomoon ($\xi$).

\section{Conclusions }\label{sec:conclusions}

\noindent In this work, we have used a one dimensional latitudinal energy balance model (LEBM) to investigate the evolution of climate on an Earth-like exomoon.  We have incorporated the effects of stellar insolation, tidal heating, eclipses of the star by the planet, atmospheric cooling and heat diffusion, and investigated how the dynamical circumstances of exomoons affect the resulting climate.

We simulated four test cases to study the exomoon climates in detail, finding that exomoons in orbits retrograde to their host planet's orbit experience stronger climate oscillations than their prograde equivalents, and are around 0.1K warmer.  In the case of moons with sufficiently short orbital periods, we find that the finite thermal relaxation time of the atmosphere allows it to act as a buffer against the otherwise strong rapid temperature oscillations produced by frequent eclipses.

We then went on to carry out an extensive parameter study of the four dimensional space constituted by the planet's semi-major axis and eccentrictity ($a_p,e_p$) and the moon's semi-major axis and eccentricity relative to the planet ($a_m,e_m$).  Comparing to the habitable zone of an Earth-like planet orbiting the same star, we find that if the exomoon's orbit is circular, the exomoon habitable manifold tends to exist at lower $a_p$ thanks to the cooling effect of eclipses, and the inner edge of the habitable zone can exist at lower $a_p$ for higher $e_p$, again as eclipses keep the moon sufficiently cool.  If the exomoon's orbit is highly eccentric, the heating budget is dominated by tidal effects, and the habitable manifold is much narrower in $a_p$, and exists at greater distance from the star.  

The simplicity of LEBMs, plus their (albeit limited) ability to be augmented by new heat sources and sinks, makes them a useful tool for studying Earth-like exomoon climates.  As they are computationally inexpensive, it is straightforward to run them multiple times, whether it is to map out their behaviour in a set parameter space as done here, or to assess the habitability of detected exomoons, by using realisations of the joint posterior distribution derived from observations.  We are keen to hone and utilise this modelling technique for both of these uses.

\section*{Acknowledgments}

\noindent DF gratefully acknowledges support from STFC grant ST/J001422/1.  DMK is supported by the NASA Carl Sagan Fellowships.

\bibliographystyle{mn2e} 
\bibliography{exomoon}

\begin{thebibliography}{43}
\expandafter\ifx\csname natexlab\endcsname\relax\def\natexlab#1{#1}\fi

\bibitem[{Abe {et~al}\mbox{.}(2011)Abe, Abe-Ouchi, Sleep, \& Zahnle}]{Abe2011}
Abe Y., Abe-Ouchi A., Sleep N.~H., Zahnle K.~J., 2011, Astrobiology, 11, 443

\bibitem[{Barnes \& O’Brien(2002)}]{Barnes2002}
Barnes J.~W., O’Brien D.~P., 2002, ApJ, 575, 1087

\bibitem[{Barnes {et~al}\mbox{.}(2009)Barnes, Jackson, Raymond, West, \&
  Greenberg}]{Barnes2009}
Barnes R., Jackson B., Raymond S.~N., West A.~A., Greenberg R., 2009, ApJ, 695,
  1006

\bibitem[{Batalha {et~al}\mbox{.}(2013)Batalha, Rowe, Bryson, Barclay, Burke,
  Caldwell, Christiansen, Mullally, Thompson, Brown, Dupree, Fabrycky, Ford,
  Fortney, Gilliland, Isaacson, Latham, Marcy, Quinn, Ragozzine, Shporer,
  Borucki, Ciardi, Gautier, Haas, Jenkins, Koch, Lissauer, Rapin, Basri, Boss,
  Buchhave, Carter, Charbonneau, Christensen-Dalsgaard, Clarke, Cochran,
  Demory, Desert, Devore, Doyle, Esquerdo, Everett, Fressin, Geary, Girouard,
  Gould, Hall, Holman, Howard, Howell, Ibrahim, Kinemuchi, Kjeldsen, Klaus, Li,
  Lucas, Meibom, Morris, Pr\v{s}a, Quintana, Sanderfer, Sasselov, Seader,
  Smith, Steffen, Still, Stumpe, Tarter, Tenenbaum, Torres, Twicken, Uddin,
  {Van Cleve}, Walkowicz, \& Welsh}]{Batalha2013}
Batalha N.~M. {et~al.}, 2013, ApJ, 204, 24

\bibitem[{Dressing \& Charbonneau(2013)}]{Dressing2013}
Dressing C.~D., Charbonneau D., 2013, ApJ, in press

\bibitem[{Eggl {et~al}\mbox{.}(2012{\natexlab{a}})Eggl, Pilat-Lohinger, Funk,
  Georgakarakos, \& Haghighipour}]{Eggl2012a}
Eggl S., Pilat-Lohinger E., Funk B., Georgakarakos N., Haghighipour N.,
  2012{\natexlab{a}}, MNRAS, 428, 3104

\bibitem[{Eggl {et~al}\mbox{.}(2012{\natexlab{b}})Eggl, Pilat-Lohinger,
  Georgakarakos, Gyergyovits, \& Funk}]{Eggl2012}
Eggl S., Pilat-Lohinger E., Georgakarakos N., Gyergyovits M., Funk B.,
  2012{\natexlab{b}}, ApJ, 752, 74

\bibitem[{Farrell(1990)}]{Farrell1990}
Farrell B.~F., 1990, Journal of Atmospheric Sciences, 47

\bibitem[{Forgan(2012)}]{Forgan2012}
Forgan D., 2012, MNRAS, 422, 1241

\bibitem[{Heller \& Barnes(2013)}]{Heller2013}
Heller R., Barnes R., 2013, Astrobiology, 13, 18

\bibitem[{Heller, Leconte \& Barnes(2011)Heller, Leconte, \&
  Barnes}]{Heller2011}
Heller R., Leconte J., Barnes R., 2011, A\&A, 528, A27

\bibitem[{Joshi(1997)}]{Joshi1997}
Joshi M., 1997, Icarus, 129, 450

\bibitem[{Kaltenegger(2000)}]{Kaltenegger2000}
Kaltenegger L., 2000, Exploration and Utilisation of the Moon. Proceedings of
  the Fourth International Conference on Exploration and Utilisation of the
  Moon: ICEUM 4. Held 10-14 July

\bibitem[{Kaltenegger \& Selsis(2010)}]{Kaltenegger2010}
Kaltenegger L., Selsis F., 2010, EAS Publications Series, 41, 485

\bibitem[{Kane \& Gelino(2012{\natexlab{a}})}]{Kane2012}
Kane S.~R., Gelino D.~M., 2012{\natexlab{a}}, Astrobiology, 12, 940

\bibitem[{Kane \& Gelino(2012{\natexlab{b}})}]{Kane2012c}
Kane S.~R., Gelino D.~M., 2012{\natexlab{b}}, Publications of the Astronomical
  Society of the Pacific, 124, 323

\bibitem[{Kane \& Hinkel(2013)}]{Kane2013}
Kane S.~R., Hinkel N.~R., 2013, ApJ, 762, 7

\bibitem[{Kasting, Whitmire \& Reynolds(1993)Kasting, Whitmire, \&
  Reynolds}]{Kasting_et_al_93}
Kasting J., Whitmire D., Reynolds R., 1993, Icarus, 101, 108

\bibitem[{Kipping(2009{\natexlab{a}})}]{Kipping2009}
Kipping D.~M., 2009{\natexlab{a}}, MNRAS, 392, 181

\bibitem[{Kipping(2009{\natexlab{b}})}]{Kipping2009a}
Kipping D.~M., 2009{\natexlab{b}}, MNRAS, 396, 1797

\bibitem[{Kipping(2011)}]{Kipping2011}
Kipping D.~M., 2011, MNRAS, 416, no

\bibitem[{Kipping {et~al}\mbox{.}(2012)Kipping, Bakos, Buchhave, Nesvorn\'{y},
  \& Schmitt}]{Kipping2012}
Kipping D.~M., Bakos G.~A., Buchhave L., Nesvorn\'{y} D., Schmitt A., 2012,
  ApJ, 750, 115

\bibitem[{Kipping, Fossey \& Campanella(2009)Kipping, Fossey, \&
  Campanella}]{Kipping_moon}
Kipping D.~M., Fossey S.~J., Campanella G., 2009, MNRAS, 400, 398

\bibitem[{Kite, Gaidos \& Manga(2011)Kite, Gaidos, \& Manga}]{Kite2011}
Kite E.~S., Gaidos E., Manga M., 2011, ApJ, 743, 41

\bibitem[{Laskar \& Robutel(1993)}]{Laskar1993}
Laskar J., Robutel P., 1993, Nature, 361, 608

\bibitem[{Liu \& Zhong(2010)}]{Liu2010}
Liu X., Zhong S., 2010, American Geophysical Union

\bibitem[{Melosh {et~al}\mbox{.}(2004)Melosh, Ekholm, Showman, \&
  Lorenz}]{Melosh2004}
Melosh H., Ekholm A., Showman A., Lorenz R., 2004, Icarus, 168, 498

\bibitem[{Parkinson {et~al}\mbox{.}(2007)Parkinson, Liang, Hartman, Hansen,
  Tinetti, Meadows, Kirschvink, \& Yung}]{plume_enceladus}
Parkinson C.~D., Liang M.-C., Hartman H., Hansen C.~J., Tinetti G., Meadows V.,
  Kirschvink J.~L., Yung Y.~L., 2007, A\&A, 463, 353

\bibitem[{Peale, Cassen \& Reynolds(1980)Peale, Cassen, \&
  Reynolds}]{Peale1980}
Peale S., Cassen P., Reynolds R., 1980, Icarus, 43, 65

\bibitem[{Porter \& Grundy(2011)}]{porter2011}
Porter S.~B., Grundy W.~M., 2011, ApJ, 736, L14

\bibitem[{Reynolds, McKay \& Kasting(1987)Reynolds, McKay, \&
  Kasting}]{Reynolds1987}
Reynolds R.~T., McKay C.~P., Kasting J.~F., 1987, Advances in Space Research,
  7, 125

\bibitem[{Sasaki, Barnes \& O'Brien(2012)Sasaki, Barnes, \&
  O'Brien}]{Sasaki2012}
Sasaki T., Barnes J.~W., O'Brien D.~P., 2012, ApJ, 754, 51

\bibitem[{Scharf(2006)}]{Scharf2006}
Scharf C.~A., 2006, ApJ, 648, 1196

\bibitem[{Spiegel, Menou \& Scharf(2008)Spiegel, Menou, \&
  Scharf}]{Spiegel_et_al_08}
Spiegel D.~S., Menou K., Scharf C.~A., 2008, ApJ, 681, 1609

\bibitem[{Spiegel {et~al}\mbox{.}(2010)Spiegel, Raymond, Dressing, Scharf, \&
  Mitchell}]{Spiegel2010}
Spiegel D.~S., Raymond S.~N., Dressing C.~D., Scharf C.~A., Mitchell J.~L.,
  2010, ApJ, 721, 1308

\bibitem[{Stamenkovic, Spohn \& Breuer(2010)Stamenkovic, Spohn, \&
  Breuer}]{Stamenkovic2010}
Stamenkovic V., Spohn T., Breuer D., 2010, American Geophysical Union

\bibitem[{Stofan {et~al}\mbox{.}(2007)Stofan, Elachi, Lunine, Lorenz, Stiles,
  Mitchell, Ostro, Soderblom, Wood, Zebker, Wall, Janssen, Kirk, Lopes,
  Paganelli, Radebaugh, Wye, Anderson, Allison, Boehmer, Callahan, Encrenaz,
  Flamini, Francescetti, Gim, Hamilton, Hensley, Johnson, Kelleher, Muhleman,
  Paillou, Picardi, Posa, Roth, Seu, Shaffer, Vetrella, \& West}]{lake_titan}
Stofan E.~R. {et~al.}, 2007, Nature, 445, 61

\bibitem[{Vladilo {et~al}\mbox{.}(2013)Vladilo, Murante, Silva, Provenzale,
  Ferri, \& Ragazzini}]{Vladilo2013}
Vladilo G., Murante G., Silva L., Provenzale A., Ferri G., Ragazzini G., 2013,
  ApJ, 767, 65

\bibitem[{Weidner \& Horne(2010)}]{Weidner2010}
Weidner C., Horne K., 2010, Astronomy and Astrophysics, 521, A76

\bibitem[{Williams \& Kasting(1997)}]{Williams1997a}
Williams D., Kasting J.~F., 1997, Icarus, 129, 254

\bibitem[{Williams(2013)}]{Williams2013}
Williams D.~M., 2013, Astrobiology, in press

\bibitem[{Williams, Kasting \& Wade(1997)Williams, Kasting, \&
  Wade}]{Williams1997b}
Williams D.~M., Kasting J.~F., Wade R.~A., 1997, Nature, 385, 234

\bibitem[{Williams \& Pollard(2002)}]{Williams2002}
Williams D.~M., Pollard D., 2002, International Journal of Astrobiology, 1

\end{thebibliography}

\appendix

\label{lastpage}

\end{document}